\documentclass[12pt, twocolumn]{article}
\usepackage[utf8x]{inputenc}
\usepackage{amsmath, amssymb, bm}
\usepackage{xcolor}
\usepackage[margin=2cm]{geometry}
\usepackage{widetext}
\usepackage{titlesec}
\usepackage{hyperref}
\usepackage{tikz}
\usetikzlibrary{positioning, hobby}

\titleformat*{\section}{\large\bfseries}
\titleformat*{\subsection}{\normalsize\bfseries}

\def\rprime{\mathbf{r}^{\prime}}

\def\hatr{\mathbf{\hat{r}}}
\renewcommand{\vec}{\mathbf}

\usepackage{calc}
\newsavebox\CBox
\newcommand\Ibar[1][0.5pt]{%
  \ifmmode\sbox\CBox{$I$}\else\sbox\CBox{I}\fi%
  \makebox[0pt][l]{\usebox\CBox}%
  \rule[0.5\ht\CBox-#1/2]{\wd\CBox}{#1}}

\title{\bf An introduction to gravitational waves through electrodynamics: a quadrupole comparison}
\author{Glauber C. Dorsch and Lucas E. A. Porto\\[3mm]
\small \it Universidade Federal de Minas Gerais, 31270-901, Belo Horizonte, MG, Brazil}
\date{}

\begin{document}

\twocolumn[
	\vspace*{-5mm}
   \begin{flushright}\textsc{Submitted to European Journal of Physics}\end{flushright}
    \maketitle
  \begin{@twocolumnfalse}
    \maketitle
    \abstract{We present a pedagogical introduction to some key computations in gravitational waves via a side-by-side comparison with the quadrupole contribution of electromagnetic radiation. Subtleties involving gauge choices and projections over transverse modes in the tensorial theory are made clearer by direct analogy with the vectorial counterpart. The power emitted by the quadrupole moment in both theories is computed, and the similarities as well as the origins of eventual discrepancies are discussed. Finally, we analyze the stability of bound systems under radiation emission, and discuss how the strength of the interactions can be established this way. We use the results to impose an anthropic bound on Newton's constant of order $G\lesssim 3\times 10^4\, G_\text{obs}$, which is on par with similar constraints from stellar formation.}
  \end{@twocolumnfalse}
  \vskip1cm
  ]

\section{Introduction}

The recent direct detections of gravitational waves (GWs) by the LIGO/Virgo collaboration~\cite{Abbott:2016blz, LIGOScientific:2018mvr, Abbott:2020niy} marks the inauguration of a new era in observational astrophysics and cosmology\footnote{For a pedagogical introduction to these results see refs.~\cite{Abbott:2016bqf, Mathur:2016cox}.}. For the first time in human history we are able not only to ``see'' distant objects in space and time through their emitted light, but also to ``hear'' the oscillations in spacetime generated by their dynamics. It is as if a new sense of perception has been opened to us, and it is nearly impossible to overestimate the impact of this discovery for our future investigations of the structure of the cosmos. 

On the one hand, the detection of GWs completes the toolkit of multi-messenger astrophysics, as we are now able to use all four known fundamental interactions as probes for astrophysical sources in the sky~\cite{Meszaros:2019xej}. The simultaneous detection of multi-messengers from the same source yields complementary information not only on the source's structure, but on the fundamental interactions themselves, especially considering that the detectable non-electromagnetic radiation typically originates from extreme conditions, such as in high-energy-density objects or under strong gravitational fields~\cite{Baker:2019nct}. Some events could not even be seen with any messenger other than GWs, e.g. black hole mergers, which are expected to emit faint (and still undetected) signals of photons and neutrinos~\cite{Meszaros:2019xej}. Moreover, from a cosmological perspective, the detection of primordial gravitational waves would yield a direct image of the Universe in its infancy, at times much earlier than we can probe nowadays with electromagnetic radiation. Indeed, the very early Universe was opaque to light, because photons were constantly being scattered by free electrons in the primordial plasma, and could not propagate freely.
Only when the Universe was about $380,000$ years old were the photons of the plasma no longer able to re-ionize the bound system of electron and nuclei, so that neutral atoms start to form and the photons become free to propagate towards us --- the Universe becomes translucid. These photons constitute the so-called cosmic microwave background radiation (CMB). 
But, because gravity is a much weaker interaction, gravitons decouple from the plasma much earlier --- possibly around the Planck epoch, at $10^{-43}$~s ---, so that the Universe has essentially always been translucid to gravitational waves. This is particularly important in light of the planned launch of LISA for 2034~\cite{Audley:2017drz}, a space-based interferometer with optimal sensitivity in the frequency range of $0.1-100$~mHz, much lower than those achievable at LIGO/Virgo on the ground. In this band we expect to detect gravitational wave relics from the electroweak phase transition~\cite{Caprini:2019egz}, which happened at the first picosecond ($10^{-12}$~s) in the history of the cosmos, so their detection would give us an image of the Universe at much earlier times than we can access with the CMB.

For all these remarkable reasons, the physics of gravitational waves has been in the spotlight for the past few years, and will remain attracting much of the attention of the community in the decades to come, since numerous open questions on detection prospects of specific sources and model constraining from data analysis still await investigation.

It is therefore of paramount importance that students be introduced to this topic already in their undergraduate years, allowing them to keep up with up-to-date discussions on one of the most relevant aspects of contemporary physics. There exist some works in the literature with exactly this introductory purpose, approaching the subject from a post-Newtonian perspective~\cite{Schutz:1984nf} or by analogy with electromagnetism via a toy ``vectorial gravitational field''~\cite{Hilborn:2017liy}. Either way, the idea is typically to obtain reasonable order of magnitude results without dealing with the intricacies of tensor calculus required for the full general relativistic computations. 

In this paper, our goal is instead to exploit the parallels between gravity and electromagnetism as a way to discuss the full tensorial calculations in a friendly way. Many results can be directly compared and translated from one theory to the other, and numerous apparent complications from the tensor nature of General Relativity can be enlightened by a direct comparison with the analog vector case in electrodynamics. Gravity and electromagnetism are the dominant interactions in our everyday experience, and much can be apprehended on the peculiarities of the fundamental interactions from a direct comparison of their similarities and their discrepancies. 

We will start in section~\ref{sec:E_noncov} with a discussion of electromagnetic radiation in a non-covariant notation. We first recapitulate some essentials of Maxwell's equations and the electromagnetic potentials in section~\ref{sec:essentials}, which should be familiar to readers in their late years of an undergraduate course, but perhaps not to their younger colleagues. The discussion is aimed to be accessible to anyone who completed a course on vector calculus, and we include it here in the interest of readability to students at the earliest possible stage of their careers. Unfortunately we cannot dispense with vector calculus, since our aim is precisely to use this as a basis for understanding some of the tensorial manipulations. For an approach requiring even less mathematical background we recommend reference~\cite{Schutz:1984nf}. In section~\ref{sec:multipole} we perform the multipole expansion, highlighting the electric quadrupole term, which, as we will see, is the analog of the leading order contribution in the gravitational case. In section~\ref{sec:power_noncov} we calculate the power emitted by this quadrupole moment, still in a non-covariant approach. Since the quadrupole is a next-to-leading-order contribution in electrodynamics, this calculation is not typically performed in introductory classes, and is even more difficult to be found in a side-by-side comparison to the gravitational case. This is one of the main contributions of the present paper. In section~\ref{sec:power_cov} we present electrodynamics in covariant notation, i.e. in a unified spacetime description appropriate in a reltivistic framework. Some subtleties regarding gauge choice and projection onto a specific gauge is discussed, which will be useful in understanding similar procedures in the gravitational case. Gravitational waves are introduced in section~\ref{sec:grav}. In section~\ref{sec:grav_quad} we perform a multipole expansion and show that the leading contribution comes from the quadrupole moment, and section~\ref{sec:grav_power} is devoted to computing the radiated power. Similarities and differences of the two cases are discussed in section~\ref{sec:comparison}, pinpointing the discrepancies to the tensorial character of the gravitational potential versus the vectorial analog of electrodynamics --- quantum mechanically we would say that the differences stem from the photon being a spin 1 particle, whereas the graviton is a spin 2. Lastly, we use the previous results to discuss in section~\ref{sec:binary} the stability of bound systems and compute an anthropic bound on Newton's constant, based on the requirement that solar systems similar to ours must be sufficiently stable under emission of gravitational radiation to allow for life formation. Our derived upper limit of $G\lesssim 3\times 10^4 \,G_\text{obs}$ is on par with similar constraints from considerations of stellar formation. Our conclusions are contained in section~\ref{sec:conclusions}. For readers unfamiliar with tensors, we present a brief discussion in appendix~\ref{sec:whytensor}, focusing on the quadrupole tensor, aiming at providing an intuition of such objects and how they relate to the more familiar vectors. In appendix~\ref{sec:groups} we discuss why tensors are important due to their peculiar transformation properties under certain symmetries, which allow us to write covariant physical laws agreed upon by all observers.

    \section{Electromagnetic radiation: non-covariant formulation}
    \label{sec:E_noncov}
    
    We start with a discussion of electrodynamics written in a non-covariant form. The aim of this section is to introduce a solid common ground on which anyone with a knowledge of vector calculus can have a firm basis, upon which the following discussions can be elaborated. Because the mathematical requirements for dealing with Maxwell's equations are much less demanding than those involved in Einstein's gravitational field equations, we use this section to discuss essentialities on the wave equation and the potential formulation of a field theory. When it comes to the gravitational case, we will make the proper analogies, but will skip a thorough deduction of the wave equation from scratch, to avoid introducing mathematical tools which are alien to our main focus. In fact, one of our main goals in this paper is to show that, once the wave equations are written down, they can be dealt with in essentially the same manner, although some important differences will be highlighted on the way.
    
    \subsection{Essentials: Maxwell's equations and potentials}
    \label{sec:essentials}
    
    The natural starting point of any discussion on electromagnetic radiation is, of course,  Maxwell's equations for the electric field $\vec{E}$ and the magnetic field $\vec{B}$. Here we will choose to work in Gaussian units and set $c=1$, since it makes the comparison with the gravitational case more transparent. In this unit system the field equations read
    \begin{equation}
    \begin{alignedat}{2}
        \nabla\cdot\vec{E} &= 4\pi\rho &&\text{(Gauss' law)}, \\ \nabla\cdot\vec{B}&=0 &&\text{(no magnetic monopoles)},\\
        \nabla\times \vec{E} &= -\frac{\partial\vec{B}}{\partial t} &&\text{(Faraday's law)},\\
        \nabla\times\vec{B} &= 4\pi\vec{J} + \frac{\partial\vec{E}}{\partial t} \quad&&\text{(Amp\`ere-Maxwell's law)},
    \end{alignedat}
    \label{eq:Maxwell}
    \end{equation}
    where $\rho$ is the electric charge density and $\vec{J}$ the associated current. 
    
    These equations already contain the statement that electric charge is conserved. Indeed, noticing that the divergence of a curl vanishes, we can take the divergence of Amp\`ere-Maxwell's equation, and use Gauss's law, to obtain the \emph{continuity equation}
    \begin{equation}
        \frac{\partial\rho}{\partial t} + \nabla\cdot\vec{J} = 0,
        \label{eq:continuity}
    \end{equation}
    which we will often use in the following developments.
    
    The absence of magnetic monopoles means that $\vec{B}$ can itself be written as the curl of some vector field,
        \begin{equation}
            \vec B = \nabla \times \vec A.
             \label{eq:B_pot}
        \end{equation}
        Plugging this identity into Faraday's law, one finds that $\vec{E}+\partial\vec{A}/\partial t$ has vanishing curl, and therefore can be written as the gradient of some scalar function $\phi$, i.e.
        
        \begin{equation}
            \vec E = -\nabla \phi - \frac{\partial \vec A}{\partial t}.
            \label{eq:E_pot}
        \end{equation}
        The fields $\phi$ and $\vec{A}$ are respectively called the scalar and vector electromagnetic potentials. Since all the information on the electromagnetic fields can be retrieved from these four quantities ($\phi$ and the three components of $\vec{A}$), it is clear that not all of the six field components are independent. By working with potentials rather than fields, we reduce the ambiguities in the description. 
        
        But that is not the end of the story. The potentials are  constructed such that the observable electric and magnetic fields can be retrieved via equations~(\ref{eq:B_pot}) and (\ref{eq:E_pot}). This prescription is not univocally defined, though. For any $\phi$ and $\vec{A}$ satisfying those equations above, we can define new potentials
        \begin{equation}
        \begin{split}
            \vec {A'} &= \vec A + \nabla \Lambda\\
            \phi^\prime &= \phi - \frac{\partial \Lambda}{\partial t}
            \label{eq:gauge}
            \end{split}
        \end{equation}
        which describe the same field configuration as before, for any scalar field $\Lambda$. This so-called \emph{gauge freedom} can be used to eliminate one additional component of the potentials, so that any electromagnetic field configuration can be described by at most three degrees of freedom\footnote{For freely propagating waves in vacuum, one extra component of the potentials can be eliminated by a residual gauge symmetry, leaving only two independent degrees of freedom. This makes explicit the fact that an electromagnetic wave in vacuum has only two polarization states. Note, however, that inside a wave guide a longitudinal component may also exist, and then three degrees of freedom are needed for describing the system.}.
        
        Writing the fields in terms of potentials, as in equations~(\ref{eq:B_pot}) and (\ref{eq:E_pot}), is enough to ensure that the two homogeneous Maxwell equations (those not involving sources $\rho$ and $\vec{J}$) will be automatically satisfied. As for the non-homogeneous equations, they acquire particularly interesting forms when we choose to work in the so-called Lorentz gauge,
        \begin{equation}
            \nabla \cdot \vec A = -\dfrac{\partial \phi}{\partial t}\quad\text{(Lorentz gauge)}.
            \label{eq:Lorentz}
        \end{equation}
        This choice can always be made, i.e. for any $\phi$ and $\vec{A}$ there will always be a function $\Lambda$ such that the transformed fields, obtained from the former by equation~(\ref{eq:gauge}), satisfy the Lorentz condition above~\cite{Griffiths:1492149}. Under these circumstances, the non-homogeneous Maxwell's equations become
        \begin{equation}
           \begin{split} 
           \frac{\partial^2 \vec{A}}{\partial t^2}  - \nabla^2\vec{A}
           &= 4\pi\vec{J},\\
          \frac{\partial^2 \phi}{\partial t^2} 
          - \nabla^2\phi 
           &= 4\pi\rho.
           \label{eq:pot_waves}
        \end{split}
        \end{equation}
        Thus, the potentials satisfy a wave equation: there exist electromagnetic waves. Let us next investigate how these waves are produced by an arbitrary source.
        
        \subsection{Multipole expansion}
        \label{sec:multipole}
        
\begin{figure}
        \centering
        \begin{tikzpicture}
            \node[inner sep=0pt, label=O] (O) at (-.7,0.2){}; \draw[fill=black] (O) circle (2pt);
            \draw[thick] (-1,-1) to[closed, curve through={(-1.2,0) .. (-1,1) .. (0,.6) .. (.8,0.25) .. (.5,-0.4) .. (.4,-.75) .. (-.6,-.7)}] (-1,-1);
            \node[inner sep=0] (rp) at (.25,-.4){};
            \node (r) at (3,.6){};
            \draw[-latex, thick] (O) -> (rp) node[pos=.3, yshift=-7pt]{$\rprime$};
            \draw[-latex, thick] (O) -> (r) node[pos=.6, yshift=5pt]{$\vec{r}$};
            \draw[-latex, thick] ([xshift=-2pt, yshift=1pt]rp.center) -> (r) node[pos=.7, yshift=-8pt]{$\vec{r}-\rprime$};
        \end{tikzpicture}
        \caption{An arbitrary source and a coordinate system with origin within it. Also shown are the vectors $\vec{r}^\prime$ pointing to a charge element in the source, $\vec{r}$ pointing to the observation point, and $\vec{r}-\vec{r}^\prime$ pointing from the element of charge to this observation point.}
        \label{fig:source}
        \end{figure}
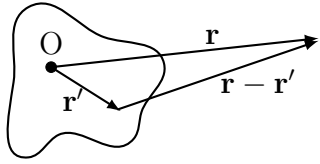

        Let us consider a localized source, and place a coordinate system with origin inside it, as in figure~\ref{fig:source}. In the static case, with no time dependence in the source and the potentials, the solution of equation~(\ref{eq:pot_waves}) at a point $\vec{r}$ has the form of a Coulomb potential $\sim (\text{source at }\rprime)/|\vec{r}-\rprime|$. Once we introduce a time variation in the source, the solution remains of a similar form, but one needs to account for the time it takes for an information at $\rprime$ to arrive at $\vec{r}$ due to the finiteness of the speed of light, as dictated by the wave equations~(\ref{eq:pot_waves}). The potential $\vec{A}(\vec{r},t)$ is actually determined by the source configuration at some point $\rprime$ at a retarded time $t_\text{ret} \equiv t-|\vec{r}-\rprime|$,
        \begin{equation}
            \vec{A}(\vec r, t) = \int \frac{\vec{J}(\vec{r'}, t - \vert \vec r - \vec{r'} \vert)}{\vert \vec r - \vec{r'} \vert} dv',
            \label{eq:Aint}
        \end{equation}
        where the integral is over the volume occupied by the source.
        
        As we are only interested in the radiation field that propagates to regions far from the source, we can limit our analysis to the case $r' \ll r$. We can then perform a so-called \emph{multipole expansion}~\cite{Jackson:100964}, noting that
        \begin{equation}\begin{split}
            \vert \vec r - \vec{r'} \vert 
            &= \sqrt{r^{2} + r'^{2} - 2 \vec{r} \cdot \vec{r'}} \\
            &\approx r\left[1 - \hatr \cdot \frac{\vec{r'}}{r} 
                + \mathcal{O}\left(\left(r^\prime/r\right)^2\right)\right]
        \end{split}\end{equation}
        and
        \begin{equation}
            \frac{1}{\vert \vec r - \vec{r'} \vert} 
            \approx \frac{1}{r}\left[
                1 + \frac{1}{r}\hatr \cdot \vec{r'} 
                +\mathcal{O}\left(\left(r^\prime/r\right)^2\right)
            \right],
            \label{eq:taylor_modulo}
        \end{equation}
        whereas the numerator can be expanded as
        \begin{equation}\begin{split}
            \!\!\vec{J}\left( \vec{r}^{\,\prime}, t\!-\! \vert \vec r\! -\! \vec {r}^{\,\prime} \vert\right)
            &= \vec{J}(\vec{r}^{\,\prime}, t\! -\! r) + 
                \left(\hatr \cdot \vec{r}^{\,\prime}\right)
                \frac{\partial \vec{J}}{\partial t}
                \\
                &\quad+\mathcal{O}\left(\left(r^\prime/r\right)^2\right).
            \label{eq:taylor_ji}
        \end{split}\end{equation}
        
        When
        \begin{equation}
             r'^{n}\frac{\partial^{n}J^{i}}{\partial t^{n}} 
            \ll r^\prime \frac{\partial J^{i}}{\partial t} \quad \forall\, n \geq 2,
            \label{eq:approx}
        \end{equation}
        the higher order terms $ \mathcal{O}\left(\left(r^\prime/r\right)^2\right)$ become increasingly smaller, and the series can be truncated. Keeping the lowest multipole terms is typically a good approximation as long as all the emitted wavelengths are much larger than the source. Alternatively, the approximation is better when the source moves sufficiently slowly, i.e. its typical oscillation period is longer than the time it takes for the emitted wave to traverse its typical dimension. Indeed, in the particular case when the source moves with a specific frequency $\omega$, condition~(\ref{eq:approx}) is equivalent to $r' \ll 1/\omega$. 
        
        Finally, substituting \eqref{eq:taylor_modulo} and \eqref{eq:taylor_ji} into equation~(\ref{eq:Aint})  yields
        \begin{equation}\begin{split}
            \vec{A}(\vec{r},t) &= \frac{1}{r}\int \vec{J}(\vec{r'}, t - r)dv' \\
            &\quad +\frac{1}{r}\left(\frac{1}{r} + \frac{\partial}{\partial t}\right)\int (\hatr \cdot \rprime)\, \vec{J}(\vec{r'}, t-r)dv' + \ldots
            \label{eq:Amulti}
        \end{split}\end{equation}
        
        The physical interpretation of these terms will be more transparent if we rewrite them in terms of the charge density rather than the current. The first term can be rewritten with the help of the identity
        \begin{equation}\begin{split}
            J^i &= \frac{\partial}{\partial x^{\prime\,j}}(x^{\prime\,i}J^j) - x^{\prime\,i}(\nabla^\prime\cdot\vec{J}\,)\\
            &= \nabla^\prime\cdot (x^{\prime\,i} \vec{J}\,) + x^{\prime\,i}\frac{\partial \rho}{\partial t},
            \label{eq:J_dp}
        \end{split}\end{equation}
       where the continuity equation (\ref{eq:continuity}) for electric charge was used in the last step. Upon integration, the divergence can be turned into a surface integral over an arbitrary surface encompassing the charge distribution. Since the surface is outside the source, the current $\vec{J}$ vanishes there, and we are left with
       \begin{equation}\begin{split}
           \vec{A}_\text{E dipole} &= \frac{1}{r}\int \vec{J}(\rprime, t-r)dv^\prime\\
           &= \frac{1}{r}\frac{\partial}{\partial t} \int \rho(\rprime, t-r)\, \rprime\, dv^\prime \equiv \frac{\dot{\vec{d}}}{r},
           \label{eq:Edipole}
       \end{split}\end{equation}
       where the dot over a quantity denotes its time derivative. Thus the leading order in the expansion for the radiative potential comes from the electric dipole moment $\vec{d}$ of the charge distribution. This is an important and well known result, stating that the monopole does not radiate. This happens because a monopole radiation would be associated to the time variation of the total charge, which vanishes because of charge conservation. We will see that, in the gravitational case, conservation of energy-momentum will lead to vanishing monopole and dipole contributions. A hint of this behaviour can already be seen in equation (\ref{eq:Edipole}): the gravitational analog of the ``electric dipole'', obtained by interpreting $\rho$ as the mass-energy density, is equal to the position of the source's center-of-mass; the  potential involves a time derivative of this quantity, i.e. the center-of-mass momentum, which is constant in the absence of external forces; therefore the gravitational dipole configuration is static and does not radiate.
       
       The second term in equation~(\ref{eq:Amulti}) can be recast as 
        \begin{equation}
            (\hatr\cdot \vec{r}^{\,\prime})\,\vec{J} = \frac{1}{2}\left[\left(\hatr \cdot \vec{r}^{\,\prime}\right)\,\vec{J} + \left(\hatr \cdot \vec{J}\right)\,\vec{r}^{\,\prime}\right] + \frac{1}{2}(\vec{r}^{\,\prime} \times \vec{J}) \times \hatr.
             \label{eq:Mdip_Equad}
        \end{equation}
        Now $\bm{\mathcal{M}}\equiv \frac{1}{2}(\rprime\times\vec{J})$ is the magnetization (the magnetic dipole moment density) due to the current $\vec{J}$, so this term will lead to the magnetic dipole radiation. The gravitational analog is obtained by replacing $\vec{J}$ by the linear momentum of the source, so the gravitational analog of $\bm{\mathcal{M}}$ is the angular momentum of the source, which is constant. This contribution to the gravitational potential is therefore also static and does not radiate\footnote{In fact, the magnetic (electric) field generated by the total magnetic dipole $m=\int \mathcal{M}dv$ can be obtained directly from the electric (magnetic) field for the electric dipole by a suitable replacement of $d\to m$~\cite{Jackson:100964}.}.
        
        The first term in the electromagnetic radiative potential which is comparable to the gravitational case is thus the electric quadrupole moment, which we can write in index notation as
        \begin{equation}
            \small
            A^{i}_\text{E quad}(\vec{r},t) = \frac{1}{2 r}\left(\frac{1}{r} + \frac{\partial}{\partial t}\right)\hat{r}_{k}\int \big( x'^{k}J^{i} + J^{k}x'^{i} \big) dv'.
            \label{eq:quad1}
        \end{equation}
        Here we used Einstein's convention that repeated indices, appearing once as a superscript and once as a subscript, ought to be summed. This is convenient to avoid crowding the notation with summation symbols. Thus a dot product $\vec{a}\cdot \vec{b} = a_x\,b_x + a_y\,b_y + a_z\,b_z = \sum_{i=1}^3 a_i b_i \equiv a_i b^i$.
        
        We can again use a trick similar to equation~(\ref{eq:J_dp}) to rewrite this in terms of the charge density via the continuity equation, using
        \begin{equation}\begin{split}
           \small \nabla'\!\cdot\!(x'^{i}&x'^{k}\vec{J}) = \frac{\partial (x'^{i}x'^{k}J^{j})}{\partial x'^{j}}\\
            &= \delta^i_{\ j} x'^{k}J^{j} + \delta^k_{\ j} x'^{i}J^{j} + x'^{i}x'^{k}\nabla^\prime\!\cdot\!\vec{J},
        \end{split}\end{equation}
        so that
        \begin{equation}\begin{split}
            &A_\text{E quad}^{i}(\vec{r},t) = \frac{1}{2 r}\left(\frac{1}{r} + \frac{\partial}{\partial t}\right)\hat{r}_{k}\times\\
            &\quad\times\left(\oint x'^{i}x'^{k}\vec{J}\cdot d\vec{S}^\prime +  \int x'^{i}x'^{k}\frac{\partial \rho}{\partial t}dv'\right).
        \end{split}\end{equation}
        
        Again the surface integral vanishes, because the surface is outside the source. Moreover, we are free to add a term proportional to $\delta^{ik}r^{\prime\,2}$ to the second integrand, since this will be a term of the form $f(t,r)\hatr$, which corresponds to a total divergence and therefore to a gauge choice. Therefore, defining the \emph{electric quadrupole tensor} as \begin{equation}
            \mathcal{Q}_{ik}(t-r) = \int \left(x'_{i}x'_{k} - \frac{1}{3}\delta_{ik}r'^{2}\right)\rho(\vec{r'}, t - r)dv',
        \end{equation} 
        we can write the electric quadrupole potential as
        \begin{equation}
            A_\text{E quad}^{i}(\vec{r},t) = \frac{1}{2 r}\left(\frac{1}{r} + \frac{\partial}{\partial t}\right)\frac{\partial}{\partial t}\big[\hat{r}_{k}\mathcal{Q}^{ik}(t-r)\big].
            \label{eq:A_nonrad}
        \end{equation}
        
        The notation can be further simplified by defining the vector
        \begin{equation}
            Q^{i}(\vec{r}, t-r) =  \hat{r}_{k}\mathcal{Q}^{ik}(t-r),
        \end{equation}
        and also by noticing that we are interested in the field behaviour in the radiation zone, at distances much larger than the wavelengths $\omega^{-1}$. This means that $(1/r+\partial/\partial t)\vec{Q} \sim (1/r + \omega)\vec{Q}\sim \omega \vec{Q}$,
        and the time derivative term dominates over the $1/r$ in equation~(\ref{eq:A_nonrad}), so we can write
        \begin{equation}
            \vec{A}_\text{E quad}(\vec{r},t) = \frac{\ddot{\vec{Q}}(\vec{r}, t-r)}{2 r},
            \label{eq:Aquad_final}
        \end{equation}
        where we recall that each dot over a quantity denotes a derivative with respect to time.
        
        Since we assume little to no previous knowledge about tensors from our readers, we propose a more in depth discussion on this \emph{quadrupole tensor} $\mathcal{Q}_{ij}$ in appendix~\ref{sec:whytensor}.
         
        \subsection{Power emitted by the electric quadrupole moment}
        \label{sec:power_noncov}
        
        The energy flux carried by the electromagnetic field is encoded in the Poynting vector $\vec{S}=(\vec{E}\times\vec{B})/4\pi$~\cite{Griffiths:1492149}. Our first task in calculating the power emitted by the quadrupole moment is then to determine the fields from the potential in equation~(\ref{eq:Aquad_final}) above. 
        
        The magnetic field is obtained from the curl of the potential, and, exploiting the fact that the charge distribution depends on $r$ only via the combination $t-r$, so that $\nabla\rho = -\dot{\rho}\hatr$, we find 
        \begin{equation}
           \begin{split}
               \vec{B}(\vec{r},t) 
             &= \frac{1}{2r} \dddot{\vec{Q}}(\vec{r},t-r)\times\hatr.
            \end{split}
        \end{equation}
        The radiative part of the electric field obeys\footnote{The time-dependent part of the electric field away from the source can be obtained from the magnetic field via Amp\`ere-Maxwell's equation in vacuum, from which the relation above can be easily deduced.}
        \begin{equation}\begin{split}
            \vec{E}(\vec r, t) & = \vec{B}(\vec{r},t)\times \hatr\\ 
            &= \frac{1}{2r} \left(\dddot{ \vec Q}(\vec r, t-r)\times \hatr\right)\times \hatr.
         \end{split}\end{equation}
    
        Notice that $\vec{E}$ and $\vec{B}$ are mutually perpendicular, and are both orthogonal to the direction of propagation $\hatr$, as expected for electromagnetic waves.  
        Hence the Poynting vector takes the form $\vec{S} = S\,\hatr$, and the total power radiated by the source is obtained by integrating this energy flux over an arbitrarily large surface at infinity,
        \begin{equation}
            \small
        \begin{split}
            &P_\text{E quad}(t) = \oint \vec{S}\cdot \hatr\,r^2d\Omega\\
            &\quad= \frac{1}{16\pi}\oint\left| \dddot{\vec{Q}}(\vec r, t-r) \times \hatr \right|^{2} d\Omega\\
                &\quad = \frac{1}{16\pi}\oint \left(\dddot{\mathcal{Q}}_{ij}\dddot{\mathcal{Q}}^{\,ik}\hat{r}^{j}\hat{r}_{k} - \dddot{\mathcal{Q}}_{ij}\dddot{\mathcal{Q}}_{kl}\hat{r}^{i}\hat{r}^{j}\hat{r}^{k}\hat{r}^{l}\right)d\Omega.
            \label{eq:Pquad_em}
        \end{split}
        \end{equation}

        The expression above is considerably simplified if we also use the identities
        \begin{equation}
            \small
            \int \left\{\!\!\begin{array}{c}
            \hat{r}^{j}\hat{r}^{k}\\
            \hat{r}^{i}\hat{r}^{j}\hat{r}^{k}\hat{r}^{l}
            \end{array}\!\!\right\}
            d\Omega = \frac{4\pi}{15}\left\{\!\!\begin{array}{c}
            5\delta^{jk}\\[1mm]
            \delta^{ij}\delta^{kl}\! +\! \delta^{ik}\delta^{jl}\! +\! \delta^{il}\delta^{jk}
            \end{array}\!\!\right\}
            \label{eq:id_delta}
        \end{equation}
        and the fact that the electric quadrupole tensor is traceless by definition, $\mathcal{Q}^i_{\ i}=0$. So, inserting the appropriate powers of Coulomb's constant $k$ and the speed of light $c$ to return to SI units, we finally arrive at
        \begin{equation}
            P_\text{E quad} = \frac{k}{20c^5} \dddot{\mathcal{Q}}_{ij}\dddot{\mathcal{Q}}^{\,ij}.
            \label{eq:PEquad}
        \end{equation}

    \section{Electromagnetic radiation: covariant formulation}
    \label{sec:power_cov}
    
     In section~\ref{sec:E_noncov} we have written electrodynamics in a vector notation which explicitly treats the three spatial dimensions on equal footing. This notation greatly simplifies the equations, for if we would rather abdicate from the techniques of vector calculus, we would need to write eight equations for six fields, as Maxwell actually did in his original paper~\cite{Maxwell:1865zz}. But then we would have to recognize that certain physical processes, namely rotations, mix some of these six fields among themselves. A change in reference frame would require an explicit transformation of the fields, and the description of the system in these terms is attached to one specific coordinate system only. The beauty and convenience of vector calculus lies precisely in allowing us to write the field equations directly in terms of rotation invariant entities, namely the vector fields $\vec{E}$ and $\vec{B}$.
    
    Now this separation between the electric and magnetic fields is also artificial, in the same sense that they are also frame-dependent quantities. One charge configuration may be perceived as static by one observer, but will correspond to a current when seen in another frame in relative motion with respect to the first. These observers will therefore disagree on the electric and magnetic fields produced by such configuration. Thus, in analogy to the case of rotations discussed above, one expects the electric and magnetic fields to be components of a larger object, and it would be desirable to write the physical laws in terms of this entity, with which all observers agree, rather than in frame-dependent terms.
    
    This is precisely the purpose of the so-called \emph{covariant formulation} of electrodynamics. We can get a glimpse of how this can be done by noting that Faraday's and Amp\`ere-Maxwell's laws establish a relation between spatial derivatives of the electric (respectively magnetic) field and time derivative of the magnetic (respectively electric) field. It is then clear that, in order to accomplish a field unification and formulate Maxwell's equations in terms of a single object, we must also treat space and time derivatives on the same footing, unifying them in a single object as well. This spacetime derivative would play the analog role of the $\nabla$ operator of vector calculus, which allows us to write Maxwell equations in terms of the vector fields themselves rather than their components.
    In other words, we want to extend the spatial symmetry of rotations to a greater set of transformations, now also involving space and time, relating frames in relative motion to one another.

    We discuss this in some more detail in appendix~\ref{sec:groups}. The main idea is that this notion of three-dimensional space plus time is superseded by a four-dimensional spacetime. An event in spacetime is described by a four-dimensional object (called a four-vector) $x^\mu=(t,\vec{x})$, where $\mu$ runs from 0 (the ``time'' component) through 1, 2 and 3 (the ``spatial'' components). The invariant length of the interval between two events is\footnote{The invariance of $ds^2$ ensures that all observers measure the same speed of light $c=1$, which is a core principle of relativity, and a defining condition  for transformations among inertial frames.} $ds^2 = c^2 dt^2 - \vec{dx}^2\equiv \eta_{\mu\nu} dx^\mu dx^\nu$, with $\eta_{\mu\nu}$ the so-called Minkowski metric of spacetime. A spacetime derivative is defined as $\partial_\mu \equiv \partial/\partial x^\mu = (\partial/\partial t, \nabla)$. If we then arrange the scalar and vector potentials, $\phi$ and $\vec{A}$, into one four-vector $A^\mu=(\phi,\vec{A})$, then from equations~(\ref{eq:B_pot}) and (\ref{eq:E_pot}) the fields can be written as\footnote{Recall that the metric can be used to raise or lower indices as in $v_\mu = \eta_{\mu\nu}v^\nu$, see appendix~\ref{sec:groups}. Since $\eta_{\mu\nu}=\text{diag}(1,-1,-1,-1)$, raising or lowering an index equal to zero changes nothing, whereas for the other indices one picks up an extra minus sign.}
    \begin{equation}\begin{split}
        E_{i} &= \partial_0 A_{i} - \partial_i A_{0},\\
        \epsilon_{ijk}B^{k} &= \partial_i A_{j} - \partial_j A_{i},
        \label{eq:fields_index}
    \end{split}\end{equation}
        where $\epsilon^{ijk}$ is the totally anti-symmetric tensor with $\epsilon_{123}=+1$ (also called Levi-Civita tensor)\footnote{Totally anti-symmetric, here, means that
        \[
            \epsilon_{ijk} =
                \left\{\begin{array}{ll}
                +1, & \text{if $ijk$ is an even permutation of 123} \\
                -1, & \text{if $ijk$ is an odd permutation of 123} \\
                0, & \text{otherwise}.
                \end{array}\right.
            \]}.
        These two definitions can be encompassed in a single object, the so-called \emph{electromagnetic} or \emph{Faraday} tensor
        \begin{equation}
            F_{\mu \nu} = \partial_\mu A_{\nu} - \partial_\nu A_{\mu}.
            \label{eq:F}
        \end{equation}
        It is easy to see that
        \begin{equation}
            F_{\mu \nu} = 
            \begin{pmatrix}
                0 & E_{1} & E_{2} & E_{3} \\
                -E_{1} & 0 & -B_{3} & B_{2} \\
                -E_{2} & B_{3} & 0 & -B_{1} \\
                -E_{3} & -B_{2} & B_{1} & 0
            \end{pmatrix},
            \label{eq:F_components}
        \end{equation}
        and that this tensor is invariant under the gauge transformation
        \begin{equation}
            A_\mu \to A_\mu + \partial_\mu \Lambda,
            \label{eq:gauge_cov}
        \end{equation}
        which is the relativistic version of equations~(\ref{eq:gauge}).
        
        Defining the four-current $j^\nu=(\rho,\vec{J})$, the two inhomogeneous\footnote{The homogeneous equations are a direct consequence of equation~(\ref{eq:F}), i.e. of writing the fields in terms of potentials.} Maxwell's equations in~(\ref{eq:Maxwell}) can be written as
        \begin{equation}
            \partial_\mu F^{\mu\nu} = \partial_\mu \partial^\mu A^\nu - \partial_\nu \partial_\mu A^\mu = 4\pi j^\nu.
            \label{eq:Maxwell_cov}
        \end{equation}
        
        It is also easy to see that, in this notation, the Lorentz gauge shown in equation~(\ref{eq:Lorentz}) becomes
        \begin{equation}
            \partial_\mu A^\mu =0\quad\text{(Lorentz gauge)},
            \label{eq:Lorentz_cov}
        \end{equation}
        and under this condition equation~(\ref{eq:Maxwell_cov}) yields
        \begin{equation}
            \partial_\mu \partial^\mu A^\nu = 4\pi j^\nu,
            \label{eq:EMwave}
        \end{equation}
        which are the wave equations~(\ref{eq:pot_waves}). 
        
        An expansion of the solution of this wave equation in multipole moments has been performed in section~\ref{sec:multipole} for the three spatial components.  Now, we are interested in computing the power emitted by this wave, as we did in section~\ref{sec:power_noncov}, but this time we want to do it in this covariant formalism. For that purpose we must notice that, in this four-dimensional spacetime paradigm, energy and momentum are unified in a single four-momentum object $p^\mu=(E,\vec{p})$. We can then obtain the Poynting vector from the \emph{energy-momentum tensor} $T^{\mu\nu}$, defined as the $\mu$-component of four-momentum per unit of a three-dimensional volume\footnote{I.e. a hypersurface in four-dimensional spacetime.} perpendicular to the $\nu$-direction in spacetime. For instance, a three-volume perpendicular to the spatial direction $j$ has two dimensions of space, with area $\mathcal{A}$, and one dimension of time with interval $\Delta t$, so $T^{0 j} = E/(\mathcal{A} \Delta t)\equiv$ rate of energy flow per unit area through surface perpendicular to the $j$-th direction. Therefore $T^{0j}$ is the $j$-th component of the Poynting vector we want. From the discussion in the previous section we know that $T^{0j} = (\vec E \times \vec B)^{j}/(4\pi)$, and we would like to rewrite this in terms of our new field object $F^{\mu\nu}$, which can be done via\footnote{The other components of the \emph{electromagnetic stress-energy tensor} describe the energy density in the fields ($T^{00}$) and the i-th component of the stress they induce on a certain surface perpendicular to the j-th direction ($T^{ij}$). A general expression for these components is 
        \[
            T^{\mu \nu} = \frac{1}{4\pi}\left(F^{\mu \alpha}F^\nu_{\ \alpha} - \frac{1}{4}\eta^{\mu \nu}F_{\alpha \beta}F^{\alpha \beta}\right).
        \]
        }
        \begin{equation}
            T^{0j} = T^{j0} = \dfrac{1}{4\pi} F^{0\alpha}F^j_{\ \alpha}.
        \end{equation}  
        The power radiated by our localized source is
        \begin{equation}
        \begin{split}
            P &= \int T^{0j} \hat{r}_{j} r^{2} d\Omega
          \label{eq:int_power_eletromag_covariant}
        \end{split}
        \end{equation}
        and, in terms of the potentials, one has
        \begin{equation}
            T^{0j} = \frac{1}{4\pi}(\partial^0 A^k - \partial^k A^0)(\partial^j A_k - \partial_k A^{j}).
            \label{eq:T0j}
        \end{equation}
        
        Now the three spatial components of the vector potential for the quadrupole moment have already been computed, shown in equation~(\ref{eq:Aquad_final}) above. But note that the expression for $T^{0j}$ above also requires us to know the scalar potential $A^0$. So in principle one would need to redo the multipole expansion for the scalar potential as well, including higher order terms in the expansions~(\ref{eq:taylor_modulo}) and (\ref{eq:taylor_ji}), which will lead to $A^0_\text{E quad} \simeq -\hat{\vec{r}}\cdot \vec{A}_\text{E quad}$ in the radiation zone. Plugging this into equations~(\ref{eq:T0j}) and (\ref{eq:int_power_eletromag_covariant}) does lead to the correct power output computed in section~\ref{sec:power_noncov} above.
        
        But we can avoid computing the scalar potential, and calculate the emitted power only from the spatial components of the potential, using an alternative approach. Even after fixing the Lorentz gauge, there is still some residual gauge freedom left because we can perform another transformation~(\ref{eq:gauge_cov}) with $\partial_\mu \partial^\mu \Lambda =0$. This ensures that if the previous potential obeyed the Lorentz gauge, so will the new one. In vacuum, we can then use this extra freedom to set
        \begin{equation}
            A^{0}_{T} = 0\implies \nabla\cdot \vec{A}_T = 0~~\left(\parbox{3cm}{\centering Coulomb or transverse gauge}\right),
            \label{eq:Coulomb_gauge}
        \end{equation}
        and we then have
        \begin{equation}
             T^{0j} = \frac{1}{4\pi} \partial^0 A^k_{T}\, \partial^j A_k^T + \text{total derivative}.
            \label{eq:poynting_covariant}
        \end{equation}
        The total derivative term contributes nothing to the average power traversing a surface at infinity and can be neglected. 
        
        The problem now is that we cannot use equation \eqref{eq:Aquad_final} directly, since that solution does not necessarily satisfy the Coulomb gauge (also called transverse gauge). We need to find the ``transverse part'' of the solution we found in equation~\eqref{eq:Aquad_final}. For that matter, one often finds in the literature that this transverse component of a vector (or a general tensor field) can be obtained by using the operator
        \begin{equation}
            \mathcal{P}_{ij} = \delta_{ij} - r_{i}r_{j},
            \label{eq:def_projec_tensor}
        \end{equation}
        which projects tensor components onto a surface orthogonal to the radial unit vector $\hat{r}$, i.e. the direction of wave propagation. At this point it is worth emphasizing that, contrary to widespread belief, this is not strictly correct: this projected field \emph{is not guaranteed} to satisfy the transverse gauge condition in equation~(\ref{eq:Coulomb_gauge})\footnote{The divergence-free component of any vector field $\vec{A}$ can be written as
        \[ \vec{A}_T \equiv \vec{A} + \frac{1}{4\pi} \nabla\cdot \int \frac{\nabla^\prime\cdot\vec{A}}{|\vec{r}-\vec{r}^{\,\prime}|}d^3r^\prime, \] and the presence of a spatial integration in the above expression clearly shows that the truly transverse (i.e. divergence-free) component of a vector field cannot in general be obtained by simply applying a \emph{local} projection operator.}. In short, when using the projector method we are actually missing information on the transverse part of the tensor. And most importantly, the error incurred by this approach may be relevant even in the radiation zone~\cite{Frenkel:2014cra, Ashtekar:2017ydh, Ashtekar:2017wgq}. However, no information on the energy flux carried by the wave is lost in this projection method, and we will therefore use it for simplicity\footnote{As a counter example, the flux of angular momentum would not be correct if calculated using this method.}. Applying the projector operator on both sides of \eqref{eq:Aquad_final}, defining $\mathcal{Q}_{ij}^T\equiv \mathcal{P}_{ik} \mathcal{Q}^k_{\ j}$ and noticing that
        \begin{equation}\begin{split}
            \partial^j A_i^{T} = -\partial_j A_i^T &= -\frac{1}{2r}\frac{\partial}{\partial x^j}\hat{r}^{l}\ddot{\mathcal{Q}}_{il}^{T}(t-r)\\
            &= \frac{1}{2r}\hat{r}^{l}\dddot{\mathcal{Q}}_{il}^{T}(t-r),
        \end{split}\end{equation}
        we find 
        \begin{equation}
            P_\text{E quad} = \frac{1}{16 \pi} \int \dddot{\mathcal{Q}}^{ik}_{T}(t-r)\dddot{\mathcal{Q}}_{il}^{T}(t-r)\hat{r}_{k}\hat{r}^{l}d\Omega.
            \label{eq:P_cov}
        \end{equation}
        Using the fact that
        \begin{equation}
          \mathcal{Q}_{ik}^{T}\mathcal{Q}_{T}^{il} = \mathcal{Q}_{ik}\mathcal{Q}^{il} - \mathcal{Q}_{ik}\mathcal{Q}^{ml}\hat{r}^{i}\hat{r}_{m},
            \label{eq:QTQT}
        \end{equation}
        we finally recover equation~(\ref{eq:Pquad_em}) obtained in the previous section, as expected.

    \section{Gravitational radiation}
    \label{sec:grav}
        The prediction of signal propagation via oscillations of the underlying fields is not an exclusive feature of electrodynamics. It is present in any theory that satisfies relativistic causality, i.e. that does not allow instantaneous transmission of information. The wave emerges in these theories precisely as the messenger carrying information from one point to another, taking a certain amount of time to traverse said distance. Electrodynamics naturally fulfils this causality requirement, and one can say that the theory was already ``born'' relativistic once it became complete after Maxwell's introduction of the displacement current in Amp\`ere's law.
        
        The historical development of our understanding of the gravitational interaction was quite different. Newtonian gravity does not involve a propagating field mediating the interaction, and in fact it contains the rather absurd notion of an instantaneous action-at-a-distance. Newton himself was dissatisfied with this feature of the theory, but at the time he did not have the tools to overcome the difficulty and therefore had to live with it. In fact, even formulating the problem in a precise statement was beyond the knowledge available at the time. Only after the advent of special relativity did it become clear that the core of the problem consists in writing a theory for the gravitational interaction which is (at least locally) covariant under the Lorentz group, respecting the symmetries of a four-dimensional spacetime.
        
        This is achieved via Einstein's equivalence principle, which states that, locally, the effect of gravity cannot be discerned from that of inertial forces felt by an accelerated observer. If an observer who is accelerated upwards drops an object,  the object will be perceived by him/her as accelerating downwards, and all objects would ``fall'' with the same acceleration, just as it would happen in a gravitational field. In fact this observer could well say that she/he is under a gravitational field, since exactly the same physics is perceived as if under such ``interaction''. Likewise, in the presence of a gravitational field, an observer in free-fall (therefore accelerated) will observe the same physics as any other inertial observer. This then establishes an equivalence between all reference frames, even between inertial and non-inertial frames, as long as they agree to disagree on the presence of a gravitational field --- in much the same way that all inertial frames are equivalent, even if they disagree on the electric and magnetic fields they observe.
        
        Since both inertial and accelerated observers must agree on their physical description, the principle of inertia should be valid for both: a free particle should undergo a ``straight'' path in its movement. But clearly that is not what we observe in the presence of gravity or, for that matter, of a fictitious force. In this case the particles' trajectories do bend. The principle of equivalence then leads us to interpret this effect as a curvature of spacetime itself: the particles still follow the ``shortest path'' possible between two points, a so-called geodesic, but in a curved spacetime a geodesic is different than an euclidean straight line. The spacetime metric is, in general, non-Minkowskian.
        
        The gravitational field is then closely linked to the metric (and other geometrical properties such as curvature) of spacetime itself, and the program of writing a relativistic theory of gravity --- so-called General Relativity because it generalizes the principle of relativity to all observers --- then consists in writing a set of equations relating the metric field to the source of gravity, which is energy and momentum. This is far from a trivial task, and it took Einstein about a decade to arrive at the desired result. We will not discuss the details here, since it would take us too far afield as it would require a lengthy digression about geometric properties of curved spaces. A thorough discussion of this endeavour can be found in many excellent textbooks on the subject~\cite{Misner:1974qy, carroll2003spacetime, Wald:106274, Weinberg:1972kfs}. Here it shall suffice for us to say that the gravitational field can be identified with the geometric curvature of four-dimensional spacetime, which is itself obtained from a non-linear combination of up to second-order derivatives of the metric field $g_{\mu\nu}$. Hence in the presence of gravity the spacetime metric is no longer Minkowski $\eta_{\mu\nu}$, but presents deviations from this flat solution which can generally be parametrized as
        \begin{equation}
            g_{\mu \nu} = \eta_{\mu \nu} + h_{\mu \nu}.
        \end{equation}
        The field equations relate this curvature to the energy-momentum $T_{\mu\nu}$ that sources the gravitational interaction, and is therefore a set of non-linear second order differential equations for the metric.
        They acquire a somewhat simpler form when written in terms of the combination
        \begin{equation}
            \bar h_{\mu \nu} \equiv h_{\mu \nu} - \frac{1}{2}\eta_{\mu \nu}h{^\alpha_\alpha},
            \label{eq:def_hbar}
        \end{equation}
        for which one has
        \begin{widetext}\begin{equation}
            -\partial_\alpha \partial^\alpha \bar{h}_{\mu\nu} - \eta_{\mu\nu} \partial^{\alpha}\partial^{\beta}\bar{h}_{\alpha\beta}
            + \partial_\nu\partial^\alpha \bar{h}_{\alpha\mu}
            + \partial_\mu\partial^\alpha \bar{h}_{\alpha\nu} +\mathcal{O}(h^2)
            =16\pi T_{\mu\nu},
            \label{eq:GR}
        \end{equation}\end{widetext}%
        where the non-linear terms in $\mathcal{O}(h^2)$ encapsulate the self-interaction of gravity, i.e. the fact that the gravitational field itself gravitates. In many interesting situations this self-interaction is negligible compared to the gravitational field produced by other matter/radiation fields and can be neglected. This is typically the case of gravitational waves, which are typically very tiny ripples over flat spacetime, with $|h_{\mu\nu}|\ll 1$, so that the gravitational backreaction can be neglected. As an example, note that the gravitational waves from binary merges detected at LIGO/Virgo have typically $|h_{\mu\nu}|\lesssim 10^{-21}$~\cite{Abbott:2016bqf}. Thus the non-linear terms $\mathcal{O}(h^2)$ can be safely neglected in the following analyses.
        
        Equation~(\ref{eq:GR}) is the analog of Maxwell's equations for the electromagnetic potentials derived in equation~(\ref{eq:Maxwell_cov}). Just as we did in section~\ref{sec:essentials}, these can be further simplified by exploiting a gauge freedom present in the theory. Namely, just as we were free to redefine the electromagnetic potentials according to eq.~(\ref{eq:gauge}) without changing the underlying physical description, so can we also perform a change in reference frame --- an arbitrary coordinate transformation --- and the physical description should remain unchanged. This is the core principle of General Relativity, that the theory is covariant under arbitrary coordinate changes, so that every observer agrees with the physics of the system under study. As a consequence of this gauge freedom, there is a redundancy in the definition of the metric tensor, such that two configurations may describe the same physics as seen by two different observers. In other words, if we perform a coordinate change according to $x^\mu\to x^\mu + \xi^\mu$, then the metric changes as
        \begin{equation}
            h_{\mu\nu} \to h_{\mu\nu} - \partial_\mu \xi_\nu - \partial_\nu \xi_\mu
            \label{eq:gauge_h}
        \end{equation}
        (see appendix~\ref{sec:groups}).
        This is the gravitational analog of gauge invariance in electrodynamics, expressed in equation~(\ref{eq:gauge_cov}). 
        These two metrics describe the same physics, so there is a redundancy in the description and we are free to choose which of these class of metrics is the most convenient to work with. Now, we can always find a gauge transformation that takes the metric to a physically equivalent configuration such that\footnote{Given a metric $\bar{h}_{\mu\nu}$, all we have to do is solve an equation of the form $\partial_\mu \partial^\mu \xi_\nu = \partial^\mu {\bar h}_{\mu\nu}$, which always has a solution. Then the new metric obtained from $\bar{h}_{\mu\nu}$ via~(\ref{eq:gauge_h}) will satisfy~(\ref{eq:Lorentz_gw}).}~\cite{Weinberg:1972kfs, Wald:106274}
        \begin{equation}
            \partial_\mu \bar h^{\mu\nu} = 0 \quad\text{(Lorentz gauge)}.
            \label{eq:Lorentz_gw}
        \end{equation}
        This is the exact analog of the Lorentz gauge condition imposed over the electromagnetic potential in equation~(\ref{eq:Lorentz_cov}), and therefore receives the same name. With this extra condition, the field equations~(\ref{eq:GR}) become
        \begin{equation}
          \partial_\alpha\partial^\alpha\bar h_{\mu \nu} = -16\pi T_{\mu \nu},
           \label{eq:lin_einstein}
        \end{equation}
        which is a wave equation for the metric perturbations.
        
        Before moving to the multipole expansion of the solution to this equation, it is worth commenting that, similarly to what happened in electrodynamics, fixing the Lorentz gauge as in~(\ref{eq:Lorentz_gw}) does not exhaust all gauge freedom in the theory. Indeed, any other transformation as in~(\ref{eq:gauge_h}) satisfying $\partial^\mu \partial_\mu \xi_\nu=0$ will leave the metric within the subset of those obeying the Lorentz gauge. We can then use the four parameters $\xi_\mu$ to fix four values of the metric, such as~\cite{Wald:106274}
        \begin{equation}
            h_{0i}=0\quad\text{and}
            \quad
            h^{\mu}_{\ \mu}=0\quad\left(\parbox{2cm}{\centering transverse traceless gauge}\right).
        \end{equation}
        This is the so-called \emph{transverse traceless} or TT gauge or also radiation gauge, and is analogous to the Coulomb gauge we explored in the treatment of electromagnetic waves.
        
        \subsection{Quadrupole moment: the leading-order contribution}
        \label{sec:grav_quad}
        
        We can now repeat the steps performed in section~\ref{sec:multipole}. To leading order, the solution to these wave equations at a point far away from the source, for the case of a sufficiently slow moving source respecting conditions analogous to~(\ref{eq:approx}), has the form
        \begin{equation}
            \bar h_{\mu\nu}(\vec r, t) = -\frac{4}{r}\int T_{\mu\nu}(\vec{r'}, t-r)dv'.
            \label{eq:int_h}    
        \end{equation}
        
        We can exploit energy-momentum conservation $\partial_\alpha T^{\nu\alpha}=0$, which follows directly from \eqref{eq:Lorentz_gw} and \eqref{eq:lin_einstein}, to write
        \begin{equation}
            T^{\mu\nu} = 
            \frac{\partial}{\partial x'^\alpha}(x'^{\mu}T^{\alpha\nu})
            = \frac{\partial}{\partial t} (x'^\mu T^{0\nu}) + \frac{\partial}{\partial x'^j}(x'^\mu T^{j\nu}),
        \end{equation}
        which is analogous to equation~(\ref{eq:J_dp}) in the electromagnetic case. As in that case, the last term will vanish upon integration over a volume encompassing the entire source. 
        The remaining term can be split into a symmetric and an anti-symmetric part, similarly to what we did before in equation~(\ref{eq:Mdip_Equad}), but now only a symmetric term is non-vanishing since the left-hand side is symmetric in the indices. Thus
        \begin{equation}
        \begin{split}
            T^{\mu\nu} &= \frac{1}{2}\frac{\partial}{\partial t} (x'^{\mu}T^{\nu 0} + x'^{\nu}T^{
            \mu 0})\\ &= \frac{1}{2}\frac{\partial}{\partial t}\frac{\partial}{\partial x'^{\alpha}}(x'^{\mu}x'^{\nu}T^{0\alpha}),
        \end{split}
        \end{equation}
        and again noticing that only the $\alpha=0$ term will survive after integration we arrive at
        \begin{equation}\begin{split}
            \bar h_{ij}(\vec r, t) &= -\frac{2}{r}\frac{\partial^2}{\partial t^2} \int x^\prime_{i}x^\prime_{j}T^{00}(\vec{r^\prime},t-r)dv'\\
            &\equiv
            -\frac{2}{r}\ddot{I}_{ij}(t-r),
            \label{eq:geracao_gw}
        \end{split}\end{equation}
        where $I_{ij}$ is the \textit{quadrupole moment tensor} of the energy-momentum distribution.
        
        We thus see that, far away from the source, the leading-order term of gravitational radiation comes from the quadrupole moment of the source, unlike the electromagnetic case where the electric dipole moment dominated. That is exactly why we dedicated some effort in section~\ref{sec:multipole} to derive properties of electromagnetic radiation associated to the electric quadrupole.
        
        Ultimately the vanishing of the monopole and dipole contributions come from energy-momentum conservation, which we have used extensively in the above derivation. Indeed, we have already remarked that a mass dipole should be something like
        $\vec R \equiv \int \vec{r'}T^{00}(r')dv'$,        which is essentially the definition of center of mass of a distribution, since $T^{00}$ corresponds to its energy density $\rho(r^\prime)$. A dipole moment radiation would scale with the second time derivative of this quantity, but conservation of momentum ensures $\ddot{\vec{R}} = 0$. Therefore, because of momentum conservation, we conclude that there can not exist radiation produced by mass dipoles. Likewise, a gravitational analog of the magnetic dipole moment would be associated to the total angular momentum of the distribution, which is also conserved and therefore has a vanishing time derivative. This is why the leading contribution to gravitational waves comes from the quadrupole.

    \subsection{Power emitted by gravitational radiation}
    \label{sec:grav_power}
    
        We have thus far tried to draw parallels between gravity and electrodynamics, but clearly both theories present differences that makes it difficult to compare them straightforwardly. While we can study electromagnetic radiation by the vector potential $\vec A$, gravitational radiation requires a quantity $h_{\mu \nu}$, that has no less than 16 components (but, to be fair, not all of them will be independent). To put both theories on a common ground we will now calculate the power emitted by a source of gravitational radiation, considering only its term associated with the quadrupole moment, obtained in the previous section. 
        
        The definition of the energy carried by gravitational waves is, in fact, a complicated matter, not only in a technical sense but also at a philosophical level. Indeed, the energy-momentum tensor of an arbitrary dynamical field is obtained from its spacetime derivatives, since momentum flow depends on how the field varies from one spacetime point to another. Now, the problem is that, in the case of the gravitational field, the metric plays the double role of the dynamical field as well as the background on which the derivatives are computed, and only the dynamical part should be associated to energy-momentum flow. But there is no clear way to separate these two parts. A promising approach consists in noticing that the gravitational wave energy-momentum is related to the backreaction non-linear terms of $\mathcal{O}(h^2)$ which we neglected in equation~(\ref{eq:GR}). We could thus rearrange the left-hand side of this equation to interpret some of these terms as a source of additional curvature --- i.e. an additional contribution to the energy-momentum tensor. The difficulty aforementioned lies precisely in attempting to split these non-linear terms in source and curvature, cause and effect, which are in fact deeply entangled to one another. A thorough discussion of this issue would take us too far afield, and we instead refer the reader to excellent discussions in refs.~\cite{Misner:1974qy, carroll2003spacetime, Wald:106274}. The bottom line is that, neglecting higher order $\mathcal{O}(h^3)$ corrections, we can define the energy-momentum flux for the gravitational wave in a consistent way by
         \begin{equation}
            T^\text{GW}_{0i} = \frac{1}{32\pi}\langle \partial_0 h^{TT}_{jk} \partial_i h^{jk}_{TT}\rangle,
            \label{eq:poynting_gw}
        \end{equation}
        where $h^{TT}_{\alpha \beta}$ denotes the perturbation $h_{\alpha \beta}$ in the transverse-traceless (TT) gauge, and $\langle \cdot \rangle$ denotes an averaging process over scales larger than a few wavelengths.
        Notice the resemblance with equation \eqref{eq:poynting_covariant}: their difference is essentially the tensor nature of the quantity $h_{jk}$ versus the vector nature of $A_{i}$, as well as an overall factor of $8$. 
        
        Note also that, since $\bar{h}_{ij}$ differs from $h_{ij}$ solely due to a factor proportional to the trace of $h_{ij}$, these two quantities coincide in the TT gauge. The power emitted from a quadrupole moment is then obtained from \eqref{eq:poynting_gw} by converting equation~\eqref{eq:geracao_gw} to the TT gauge. The task is analogous to what we encountered in section~\ref{sec:power_cov}, in that we need to extract the transverse (and traceless) components of the previous solution. This can be done via a projection tensor defined in terms of the projector already discussed in equation~\eqref{eq:def_projec_tensor} via
        \begin{equation}
            \Pi_{ij}^{\hphantom{ij}kl} \equiv \left(\mathcal{P}_i^{\  k}\mathcal{P}_j^{\ l} - \frac{1}{2}\mathcal{P}_{ij}\mathcal{P}^{kl}
            \right).
            \label{eq:def_tensorTT}
        \end{equation}
        It is easy to check that any tensor $h_{ij}^{TT}\equiv \Pi_{ij}^{\hphantom{ij}kl}h_{kl}$ indeed satisfies  $\partial^i h^{TT}_{ij} = 0$ and $h^{TT\, i}_{\hphantom{TT}\ i} = 0$.
        
        We then apply this projector to \eqref{eq:geracao_gw} to obtain an expression for the metric perturbations in terms of the TT component of the quadrupole tensor, $I_{ij}^{TT}$. But because this is traceless, it is surely equal to the transverse traceless part of the traceless quadrupole tensor
        \begin{equation}\begin{split}
            \Ibar_{ij} &\equiv I_{ij} - \frac{1}{3}\delta_{ij}\delta^{kl}I_{kl} \\
            &= \int \left(x'_{i}x'_{j} - \frac{1}{3}r^{\prime\,2}\delta_{ij}\right) T^{00}(\vec r^\prime,t)dv^\prime,
            \label{eq:Iij}
        \end{split}\end{equation}
        so 
        \begin{equation}
            h_{ij}^{TT}(\vec r, t) = -\frac{2}{r}\ddot\Ibar_{ij}^{\, TT}(t-r),
            \label{eq:h_tt}
        \end{equation}
        which can be directly compared to the electromagnetic equations~(\ref{eq:A_nonrad}) and~(\ref{eq:Aquad_final}).

        Now, substituing this into the expression \eqref{eq:poynting_gw} for the energy flux, neglecting terms that decay faster than $1/r$, and integrating over a surface at infinity, we obtain the total radiated power
        \begin{equation}
            P_\text{GW quad} = \frac{1}{8\pi}\left\langle \int \dddot{\Ibar}_{jk}^{TT} \dddot{\Ibar}^{jk}_{TT}d\Omega \right\rangle,
            \label{eq:p_J_TT}
        \end{equation}
        which is completely analogous to equation~(\ref{eq:P_cov}) in electrodynamics.
    
        Finally, using \eqref{eq:def_tensorTT} and the fact that $\Ibar^i_{\, i} = 0$, it is possible to show that
        \begin{equation}\begin{split}
            \dddot{\Ibar}_{jk}^{\, TT} \dddot{\Ibar}^{\, jk}_{TT} &= \dddot{\Ibar}_{jk} \dddot{\Ibar}^{\,jk} - 2\dddot{\Ibar}_{\ i}^j\dddot{\Ibar}^{\, ik}\hat{r}_j\hat{r}_k\\
            &\qquad+ \frac{1}{2}\dddot{\Ibar}^{\, ij}\dddot{\Ibar}^{\, kl}\hat{r}_i\hat{r}_j\hat{r}_k\hat{r}_l
            \label{eq:J_TT_from_J}
        \end{split}\end{equation}
        (compare equation~(\ref{eq:QTQT})),
        and we then have finally
        \begin{equation}
            P_\text{GW quad} = \frac{G}{5c^5}\langle \dddot{\Ibar}_{ij}\dddot{\Ibar}^{ij}\rangle.
            \label{eq:pot_gw}
        \end{equation}
        
    \section{Similarities and differences}
    \label{sec:comparison}
    Having performed these computations, let us now compare the results obtained in equations~ \eqref{eq:PEquad} and \eqref{eq:pot_gw}, which we rewrite here for convenience:
    \begin{equation}
    \begin{split}
        P_\text{E quad} &= \frac{k}{20c^5}\langle \dddot{\cal Q}_{ij}\dddot{\cal Q}^{ij} \rangle,
        \\
        P_\text{GW quad} &= \frac{G}{5c^5}\langle \dddot{\Ibar}_{ij}\dddot{\Ibar}^{ij}\rangle.
    \end{split}\end{equation}
    There is a clear similarity between the two formulas, in that one can be obtained from the other essentially by doing the rather intuitive replacements $G \longleftrightarrow k$, $\rho_\text{charge} \longleftrightarrow \rho_\text{mass}$ and, consequently, $Q_{ij} \longleftrightarrow \Ibar_{ij}$. However, one notices that, after these substitutions, the two formulas still differ by a factor of 4, more specifically $P_\text{GW quad} \longleftrightarrow 4P_\text{E quad}$. There are three sources for this discrepancy, which we henceforth discuss.
    
    \subsubsection*{Fundamental equations}
    The first thing that might produce such a discrepant factor is the difference between the two field equations, which can be put in the form of the two wave equations \eqref{eq:EMwave} and \eqref{eq:lin_einstein},
    \begin{equation}\begin{split}
        \partial_\alpha \partial^\alpha A^\mu &= 4\pi J^\mu\\
        \partial_\alpha \partial^\alpha\bar h_{\mu \nu} &= -16\pi T_{\mu \nu}.
    \end{split}\end{equation}
    We see that the numerical factor of these equations differs by a factor 4. This is the origin of the numerical factor difference in equations \eqref{eq:Aquad_final} and \eqref{eq:geracao_gw},
    \begin{equation}
    \begin{split}
         A_{i}^{T}(\vec{r},t) &= \frac{1}{2 r}\hat{r}^{j}\ddot{\cal Q}_{ij}^{T}(t-r),\\
        h_{ij}^{TT}(\vec r, t) &= -\frac{2}{r}\ddot{\Ibar}_{ij}^{\,TT}(t-r).
        \label{eq:Ah}
    \end{split}\end{equation}
    
    \subsubsection*{Energy-momentum tensor}
    The energy flux scales with the fields squared, and if we recall the definition of the energy-momentum tensor for the propagating electromagnetic and gravitational fields, given by equations \eqref{eq:poynting_covariant} and \eqref{eq:poynting_gw},
    \begin{equation}\begin{split}
        T^\text{EM}_{0j} &= \frac{1}{4\pi} \partial_0 A^{i}_{T}\partial_j A_{i}^{T},\\
        T^\text{GW}_{0j} &= \frac{1}{32\pi}\langle \partial_0 h^{TT}_{ik} \partial_j h^{ik}_{TT}\rangle,
    \end{split}\end{equation}
    one notices that the prefactor differs by $1/8$. So, putting this together with the factor difference discussed above, one would expect, based solely on these different definitions of the relevant quantities, that the GW power output should be larger by a factor 2 only. But there is one additional and very important difference between the two theories.

    \subsubsection*{Vector \emph{vs} tensor nature of the fields}
    
    The most remarkable difference between electrodynamics and gravity is the fact that the latter interaction is mediated by a field described by a rank-2 tensor (namely the metric), whereas for electrodynamics one deals with vector fields (the four-potential). In other words, the photon is a spin 1 particle, the graviton a spin 2. Ultimately, this is the reason for the additional factor $2$ found in the above analysis, since it can be shown that Maxwell's equations and the linearized version of Einstein's general relativity are the only consistent linear equations for a spin 1 and a spin 2 particle~\cite{Feynman, Deser:1969wk}. However, as we will now show, this spin difference between the mediators of these two interactions will provide yet an extra factor of 2 enhancement in the power output of GW quadrupole as compared to the electromagnetic case, simply due to the tensorial versus vectorial character of the respective potentials.
    
    To see how that comes about, recall equations~(\ref{eq:P_cov}) and (\ref{eq:p_J_TT}) for the power output in each case,
       $P_\text{E quad}\sim \int (\dddot{\mathcal{Q}}^{ik}_{T} \hat{r}_k) (\dddot{\mathcal{Q}}_{il}^{T} \hat{r}^l)\, d\Omega$ and $
       P_\text{GW quad}\sim \int \dddot{\Ibar}_{jk}^{TT} \dddot{\Ibar}^{jk}_{TT}\,d\Omega.$
    In both cases the quadrupole moment is a rank 2 tensor, as we discussed in section~\ref{sec:whytensor}, but the electromagnetic field is a vector (spin 1), so we must find a way to construct a vector out of a two-indexed object. The only consistent way is to multiply the tensor with a vector and contract one of their indices\footnote{This can be seen as a multiplication of an $n\times n$ matrix by an $n\times 1$ vector yielding a vector.}, and the only available vector that can be used for this purpose is the direction $\hat{r}$. Thus $A^T_i \sim {\cal Q}^{T}_{ij}\hat{r}^j$, as in equation~(\ref{eq:Ah}). Thus there are two factors of $\hat{r}$ appearing already in the definition of the radiated power, and two other factors will appear from the projector ${\cal P}_{ij}\equiv \delta_{ij} - \hat{r}_i \hat{r}_j$ used to obtain the transverse component of the quadrupole, ${\cal Q}^T_{ij}$, from the full tensor ${\cal Q}_{ij}$. The procedure leads to equation~(\ref{eq:QTQT}) which, upon integration, yields $P_\text{E quad} \sim Q_{ij}Q^{ij}/5$, apart from prefactors already addressed in the discussion above. On the other hand, the metric is itself a rank-2 tensor like the quadrupole moment, so one can have (and one indeed does have) $h^{TT}_{ij}\sim \Ibar^{TT}_{ij}$, as in~equation~(\ref{eq:Ah}). There are no factors of $\hat{r}$ in the integrand for the total power when it is written in terms of the transverse traceless quadrupole moments. But once we go back to the full quadrupole tensor, because of the tensorial character of the metric, the associated projector is slightly more complex than for the vector case, cf.~equation~(\ref{eq:def_tensorTT}), so that the power is given by the integral of terms as in equation~(\ref{eq:J_TT_from_J}), which yields $P_\text{GW quad} \sim 2 \Ibar_{ij} \Ibar^{ij}/5$.
    
    We thus see that the mere tensorial nature of gravity, or the spin 2 of the graviton, in comparison to the spin 1 photon, is the root of an additional factor of $2$ enhancement in the radiated power. 
    
    Combining this with the other factor $2$ discussed above we find a discrepancy by an overall factor of $4$, in agreement with our explicit calculations.

    \section{Binary systems and an anthropic bound on $G$}
    \label{sec:binary}
    
        Finally, we can use these results at hand, concerning the power emitted by electromagnetic and gravitational waves, to discuss the hierarchy among these two interactions in terms of the stability of their bound systems.
        
        Let us consider two objects of masses $M\gg m$ orbiting one another with angular frequency $\omega$, and for simplicity assume that the heavier one stays still at the center of mass. 
        Two instances of such model would be the Earth-Sun system, and also an electron orbiting the nucleus. In both cases one can check that the ``slowly moving'' condition in equation~(\ref{eq:approx}) holds\footnote{The gravitational waves take about $8$ minutes to cross from the Sun to the Earth, much smaller than the one year period of revolution of the Earth around the Sun. As for the electron in an atom, one can use Bohr's model to estimate an angular frequency $\omega \sim \hbar/(m_e R^2) \sim  10^{15}~\text{s}^{-1}$, where $\hbar$ is the reduced Planck constant, $m_e$ is the electron mass and $R\sim 5\times 10^{-11}$~m the atomic (Bohr) radius. Then clearly $R/c \sim 10^{-19}~\text{s}\ll 1/\omega$.}, so we can truncate the multipole expansion at some desired order. The energy density in these non-relativistic systems is dominated by the rest masses, and is given simply by
        \begin{equation}\begin{split}
            T^{00} &= M\delta(x^1)\delta(x^2)\delta(x^{3})\\
            &~~ + m\delta(x^{3})\delta(x^{1}\! -\! R\cos\omega t)\delta(x^{2}\! -\! R\sin\omega t),
        \end{split}\end{equation}
        where we assumed a circular motion of radius $R$ for simplicity\footnote{The case of elliptical orbits has been discussed in ref.~\cite{PhysRev.131.435}, where it has been shown that the energy loss increases with the eccentricity. This is expected from the fact that, the larger the eccentricity, the faster the object will move  as it approaches the perihelion. However, for the Earth-Sun orbit the eccentricity is $\sim 0.02$ and the total power would be enhanced by only $\sim 0.2\%$ compared to our estimates. }. 
        A straightforwad calculation using the definition in~(\ref{eq:Iij}) and the result in~(\ref{eq:h_tt}) yields the transverse-traceless metric 
        \begin{equation}
            h^{TT}_{i j} = \frac{4mR^2\omega^2}{r}
            \begin{pmatrix}
                \cos(2\omega t) & \sin(2\omega t) & 0 \\
                \sin(2\omega t) & -\cos(2\omega t) & 0 \\
                0 & 0 & 0   \\
            \end{pmatrix},
        \end{equation}
        and using equation~(\ref{eq:pot_gw}) we find the radiated power in gravitational waves
        \begin{equation}
            P_\text{GW} = \frac{32}{5}\frac{Gm^2 R^4\omega^6}{c^5}.
            \label{eq:P_system}
        \end{equation}
        Note that the metric field, and therefore the gravitational wave, oscillates with twice the frequency of the source. This is a typical behaviour of the quadrupole radiation. We discuss this in appendix~\ref{sec:whytensor}.
        
        Now, since the system is slowly moving, non-relativistic mechanics works just fine. So, focusing first on the case of an electron in a (classical model of the) hydrogen atom\footnote{We extrapolate some aspects of classical mechanics to the case of the electron in an atom, in the spirit of Rutherford's model.}, and noticing that the dominant centripetal force is the Coulomb attraction between electron and nucleus, we find
        \begin{equation}
            m\omega^2R = \frac{k_e e^2}{R^2},
            \label{eq:ac_ctp}
        \end{equation}
        with $k_e$ the Coulomb constant. The total kinetic + potential energy (neglecting rest masses, which will remain unchanged throughout the entire motion) is $E = m\omega^2 R^2/2 - k_e e^2/R = - k_e e^2/(2R)$. This energy is reduced upon emission of (gravitational and electromagnetic) waves, which will tend to make the orbit collapse. To find the lifetime of the system taking into account only emission of GWs we can equate $dE/dt = -P_\text{GW}$, which gives
        \begin{equation}
            R^3\frac{dR}{dt} = - \frac{64}{5}\frac{(Gm^2)(k_e e^2)^2}{ m^3 c^5}.
        \end{equation}
        
        This can be integrated to find the time it takes for the system to collapse from the initial radius $R_0$  down to zero, to find the so-called \emph{spiral time}
        \begin{equation}
            \tau^\text{atom}_\text{GW} = \frac{5}{256} \frac{m^3 c^5 R_0^4}{(Gm^2)(k_e e^2)^2} \sim 10^{35}~\text{s}.
            \label{eq:spiralGW}
        \end{equation}
        This is 18 orders of magnitude larger than the age of the Universe! Therefore the classical atom (i.e. the Rutherford model) is completely stable under the emission of gravitational waves. However, we can likewise obtain the lifetime of the classical hydrogen atom considering also the emission of electromagnetic waves. If we ignore the dipole contribution for a moment, then the emitted power can be directly obtained from~(\ref{eq:P_system}) via the replacement $Gm^2 \to k_e e^2/4$ (the factor of 4 is discussed in section~\ref{sec:comparison} above) and one finds a lifetime $\tau_\text{E quad}\sim  10^{-6}~\text{s}$. There is thus a discrepancy of 42 orders of magnitude, due to the fact that the electromagnetic interaction is much stronger than gravity\footnote{This fact is known as the \emph{gauge hierarchy} in particle physics, and its origin is still a great mystery, being at the root of the so-called \emph{hierarchy problem}.}$^,$\footnote{The discrepancy is actually aggravated by the fact that the leading contribution to electromagnetic radiation comes from the dipole moment. In this case one can show, using the same machinery developed above, that $P_\text{E dipole} = 2 k_e |\ddot{\vec{d}}|/3c^3$, with $\vec{d}$ is the dipole moment of the distribution, and the lifetime of the classical atom would then be $\tau_\text{E dipole}\sim 10^{-11}$~s. The existence of electromagnetic waves is therefore one strong motivation for abandoning the classical model of the atom, ultimately leading to a quantum mechanical description. }. On general grounds one can say that the lifetime of a gravitationally bound system is typically much larger than an electromagnetically bound one. 
    
       The calculations above can be directly translated to the Earth-Sun system via the replacements $k_e e^2\to GMm$ in equation~(\ref{eq:ac_ctp}) so that
       \begin{equation}\begin{split}
           \tau_\text{Earth-Sun} &= \frac{5}{256} \frac{c^5 R_0^4}{G^3M^2 m} \\
           &\approx (1.2\times 10^{23}~\text{years}) \left(\frac{G_\text{obs}}{G}\right)^3,
       \end{split}\end{equation}
       with $G_\text{obs}= 6.674\times 10^{-11}~\text{N}\, \text{m}^{2}\,\text{kg}^{-2}$ the currently measured value of Newton's constant~\cite{CODATA}. Again, the lifetime of this bound system due to emission of gravitational waves is many orders of magnitude larger than the age of the Universe, so the system is completely stable, as should be expected. However, the scaling of this lifetime with $G^{-3}$ shows that, if Newton's constant were somewhat larger than the observed value, then the solar system would have collapsed before it reached its current age, and before intelligent life would have formed. This places an anthropic upper bound on Newton's constant of
       $
           G\lesssim 3\times 10^4 \, G_\text{obs}
       $, which is similar to other anthropic bounds on $G$ obtained from conditions on stellar formation~\cite{Adams:2008ad, Barnes:2011zh}. The latter depend, however, on other parameters associated to nuclear reaction rates, whereas here we find a direct bound on $G$ alone.

    \section{Conclusions}
    \label{sec:conclusions}
    
    The recent detection of gravitational waves opened a new venue for exploring the workings of our cosmos, and it will certainly have an impact on many fields of research such as astrophysics, cosmology and even particle physics. It is therefore of utmost importance that undergraduate students become acquainted with the basic tools to understand, discuss and eventually also to work on the subject.
    
    In this paper we have presented some aspects of gravitational wave physics by exploring parallelisms with the often more familiar theory of electrodynamics.    One important difference\footnote{Another important difference is the fact that gravity is self-interacting whereas the electromagnetic field is not, because it does not carry electric charge. This is why Maxwell's equations are linear and the field equations in general relativity are not. However, this difference is not so important in discussions of gravitational waves, where the linear regime can be assumed because backreactions are negligible.} between the two theories is the fact that electrodynamics can be described by a four-vector, whereas for gravity the fundamental object is a rank-2 tensor. Put another way, the graviton (the particle mediating the gravitational interaction) is a spin 2 particle, whereas the photon (mediator of electromagnetism) has spin 1. It then follows~\cite{Couch:1972wh} that electromagnetic radiation is predominantly produced by the dipole moment of the source, whereas the leading order contribution to gravitational waves stems from the quadrupole. However, there are also electromagnetic waves produced by quadrupole moments, even if usually subleading, and the comparison with the gravitational case is then possible: in both cases the potential is directly proportional to the second derivative of the quadrupole moment tensor of the form $\int (x_i x_j - \delta_{ij}r^2/3)\rho\, d^3x$, where $\rho$ is the density of the charge of the interaction (i.e. electric charge or mass).
    
    This side-by-side display of the main calculations leading to the quadrupole radiation in both interactions is an important contribution of the present work. It allows many early career physicists to have a first introduction to the subject of gravitational waves through the more familiar topic of electrodynamics, while at the same time learning about both through a direct comparison of their similarities and differences.
    
    We have also studied the emitted power by the quadrupole radiation in both cases, finding a relation of the form $P\sim (\text{coupling constant})\times (\text{quadrupole tensor})^2$. One could na\"ively expect that, once we know the power radiated via one of these interactions, one could find the other case via the immediate replacement of (Newton's constant) $\leftrightarrow$ (Coulomb's constant) and (mass density) $\leftrightarrow$ (charge density). We have shown that this expectation is essentially correct, but not totally, due to a difference by a factor of $4$ which is ultimately due to the differing spins of the photon and the graviton. 
    
    Finally, we have used the results to study the stability of bound systems under radiative emissions. This allowed us to understand the hierarchy of the interaction strengths in terms of timelifes of bound systems\footnote{One can also compare the strength of the other fundamental interactions in the same fashion, using the lifetime of the bound systems they form, as argued e.g. in reference~\cite{Halzen:1984mc}.}, and also place an anthropic bound on Newton's gravitational constant which is on par with similar bounds from stellar formation, but with the advantage that no dependence on nuclear physics parameters are involved in our case: the bound is imposed directly on gravity only.
    
    Since we have discussed at some length gravitational waves in the context of general relativity, we should conclude by emphasizing that the detection of gravitational waves constitutes a testimonial of the relativistic nature of gravity, but not of general relativity in particular~\cite{Schutz:1984nf}. Most relativistic theories will predict radiation emission, since it is a mere consequence of retarded interactions. In fact, the detection of gravitational waves could be a new and rich alley for testing GR and probing alternative theories of gravity~\cite{TheLIGOScientific:2016src, LIGOScientific:2019fpa}.
    
    \section*{Acknowledegements}
    
    GCD would like to acknowledge the support from Pr\'o-Reitoria de Pesquisa of Universidade Federal de Minas Gerais (UFMG) under grant number 28359.
    
    \appendix
    
    \section{Why is the quadrupole moment a tensor?}
    \label{sec:whytensor}
        
        In section~\ref{sec:E_noncov} we saw that the \emph{quadrupole tensor} $\mathcal{Q}_{ij}$ emerged naturally from the mathematical development of the multipole expansion of the potential. But what kind of object is this? Why does it have two indices? And what is its physical interpretation? 
        
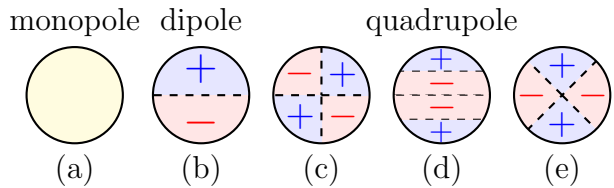
\begin{figure}
            \centering
            \def\d{-5pt}
            \def\q{-3pt}
            \begin{tikzpicture}[scale=.8]
                \node (O1) at (-2.1,0){};
                \draw[thick, fill=yellow!15] (O1) circle (8mm) node[yshift=-1cm]{(a)};
                \node[above=5mm of O1] {monopole};
                \node (O) at (0,0){};
                \begin{scope}
                    \clip (O) circle (8mm);
                    \fill[blue!10] ([xshift=-8mm]O) rectangle ([xshift=8mm, yshift=9mm]O) node[pos=.5, color=blue]{\Large $+$};
                    \draw[dashed, thick, fill=red!10] ([xshift=-8mm]O) rectangle ([xshift=8mm,yshift=-9mm]O) node[pos=.5, color=red]{\Large $-$};
                \end{scope}
                \draw[thick] (O) circle (8mm) node[yshift=-1cm]{\normalsize (b)};
                \node[above=5mm of O] {dipole};
                \node (O2) at (6,0){};
                \begin{scope}
                   \clip (O2) circle (8mm);
                    \fill[dashed, thick, rotate around={-45:(O2)}, fill=red!10] (O2) rectangle ([xshift=1cm, yshift=1cm]O2) node[pos=.36, color=red]{\large $-$};
                    \fill[dashed, thick, rotate around={135:(O2)}, fill=red!10] (O2) rectangle ([xshift=1cm, yshift=1cm]O2) node[pos=.36, color=red]{\large $-$};
                    \draw[dashed, thick, rotate around={-135:(O2)}, fill=blue!10] (O2) rectangle ([xshift=1cm, yshift=1cm]O2) node[pos=.35, color=blue]{\large $+$};
                    \draw[dashed, thick, rotate around={45:(O2)}, fill=blue!10] (O2) rectangle ([xshift=1cm, yshift=1cm]O2) node[pos=.35, color=blue]{\large $+$};
                \end{scope}
                \draw[thick] (O2) circle (8mm) node[yshift=-1cm]{\normalsize (e)};
                \node (O3) at (4,0){};
                \begin{scope}
                   \clip (O3) circle (8mm);
                    \fill[blue!10] ([yshift=4mm, xshift=8mm]O3) rectangle ([yshift=1cm,xshift=-8mm]O3) node[xshift=18pt, yshift=-9pt, color=blue]{\normalsize $+$};
                    \fill[blue!10] ([yshift=-1cm, xshift=8mm]O3) rectangle ([yshift=-4mm,xshift=-8mm]O3) node[xshift=18pt, yshift=-5pt, color=blue]{\normalsize $+$};
                    \draw[dashed,  fill=red!10] ([xshift=8mm]O3) rectangle ([yshift=4mm, xshift=-8mm]O3) node[pos=.5, color=red]{\large $-$};
                    \draw[dashed, fill=red!10] ([yshift=-4mm, xshift=8mm]O3) rectangle ([xshift=-8mm]O3) node[pos=.5, color=red]{\large $-$};
                \end{scope}
                \draw[thick] (O3) circle (8mm) node[yshift=-1cm]{\normalsize (d)};
                \node (O4) at (2,0){};
                \begin{scope}
                   \clip (O4) circle (8mm);
                    \fill[dashed, thick, rotate around={90:(O4)}, fill=red!10] (O4) rectangle ([xshift=1cm, yshift=1cm]O4) node[pos=.36, color=red]{\large $-$};
                    \fill[dashed, thick, rotate around={-90:(O4)}, fill=red!10] (O4) rectangle ([xshift=1cm, yshift=1cm]O4) node[pos=.36, color=red]{\large $-$};
                    \draw[dashed, thick, rotate around={180:(O4)}, fill=blue!10] (O4) rectangle ([xshift=1cm, yshift=1cm]O4) node[pos=.35, color=blue]{\large $+$};
                    \draw[dashed, thick, rotate around={0:(O4)}, fill=blue!10] (O4) rectangle ([xshift=1cm, yshift=1cm]O4) node[pos=.35, color=blue]{\large $+$};
                \end{scope}
                \draw[thick] (O4) circle (8mm) node[yshift=-1cm]{\normalsize (c)};
                \node[above=5mm of O3, 
                ]{quadrupole};
            \end{tikzpicture}
            \caption{Representation of the (a) monopole, (b) dipole and (c-e) quadrupole moments of a certain charge distribution. The monopole encodes the behaviour of the source as a whole, neglecting any substructure. The dipole contains information on polarization, whereas the quadrupole contains a finer-grained information on how the charge is distributed along quadrants. The images should be rotated around the vertical axis to obtain a three-dimensional distribution, and the charges are to be seen as distributed over the \emph{surface} of the sphere, not the entire volume.}
            \label{fig:poles}
        \end{figure}

        In order to gain some insight, let us step back and first investigate the other moments of the multipole expansion. At zeroth order, the most important aspect of a certain configuration is the total charge, the so-called ``monopole''. If the source is viewed from a large enough distance, the monopole  completely dominates and the source behaves as if it were point-like, i.e. all other little details about its charge distribution become irrelevant. The monopole is just a number encapsulating a property of the source \emph{as a whole}, see figure~\ref{fig:poles} (a). Clearly its value does not depend on the coordinate system we use, for instance it does not change under rotations: it is what we call a \emph{scalar} quantity.
        
        But as we approach the source we start to detect its substructure as well, being able to probe how this total charge is distributed in a certain volume. Because the monopole already encodes information on the amount of charge in the \emph{entirety} of space, the simplest additional information would come from partitioning space in two halves and evaluating the imbalance of charge distribution between these regions, as in figure~\ref{fig:poles} (b). 
        For each manner of partitioning space in halves we get a number, a product of the charge and the geometry of the source under such splitting. For instance, in the simple case of two homogeneously and oppositely charged hemispherical shells, as depicted in figure~\ref{fig:poles} (b), this number would be just the total charge in each hemisphere times their diameter. Now, there are three independent orthogonal planes that can be used for separating space in two halves, so we need three components to fully describe this \emph{dipole moment} of the distribution. And these components depend on the planes we picked in the first place, i.e. on the chosen coordinate system. This makes clear how and why the dipolar feature of the distribution is described by a vector, an object indicating a single direction in space. 
        
        Still this is too simplistic, and a finer graining may be needed to better characterize the source. The next best information consists in partitioning it in fourths and analysing how the charge is distributed along each of these regions. Let us count the ways in which we can do this. There are three possibilities of partitioning space in four quadrants by two axes, as shown in figure~\ref{fig:poles} (c), e.g. when these axes are $xy$, $xz$ and $yz$. But it is easy to see that, even if we combine these three cases, we still do not get the most general charge distribution, because the poles in these three configurations always have vanishing charge. This limitation can be remedied by adding three other partitions in quartiles, this time perpendicular to a given axis, as in figure~\ref{fig:poles} (d). Alternatively we can also rotate the partitions in figure (c) by $45^\circ$, obtaining new quadrants as in figure~\ref{fig:poles} (e). It turns out, however, that while three partitions in quadrants is not enough to fully characterize a distribution over a sphere, six different partitions is too much: only five of them are linearly independent.    This means we need only five components to fully characterize the quadrupole moment, which agrees with the counting of the number of elements in a symmetric traceless tensor such as $\mathcal{Q}_{ij}$. The  components $\mathcal{Q}_{xz}$,  $\mathcal{Q}_{yz}$ and $\mathcal{Q}_{xy}$ encode information on the charge distribution along the quadrants defined by the $xz$, $yz$ and $xy$ axes, figure (c). The partition in figure (e), where the positive and negative poles are aligned along the $y$ and $x$ axes (respectively) is associated to $\mathcal{Q}_{yy}-\mathcal{Q}_{xx}$. Finally, the axial partition with planes perpendicular to the $z$ axis, figure~(d), is encoded in $\mathcal{Q}_{zz}$.
        
        Note that, to characterize the quadrupole distribution, it is in general necessary to specify \emph{two} directions in space. Each vector defines an orthogonal plane dividing space in two regions, so we need two vectors to define four quadrants. The quadrupole moment is therefore described by a two-index structure, and we say it is a tensor of rank 2. Just as a vector can be represented by a column, so can this two-indexed object be represented by a matrix. And because $\mathcal{Q}_{ij}$ is a symmetric matrix, its eigenvectors are orthogonal\footnote{Strictly speaking this is true only for non-degenerate eigenvalues. There are situations in which only one direction will be singled out, as in an axial distribution shown in figure~\ref{fig:poles} (d). In this case there will be a doubly degenerate eigenvalue, corresponding to the freedom of choosing the axes perpendicular to the symmetry direction.} and point in the directions of the electropositive and electronegative regions. In the simple case of a distribution shown in figure~\ref{fig:poles} (e), these would be the horizontal and vertical directions, with negative and positive eigenvalues respectively.
        
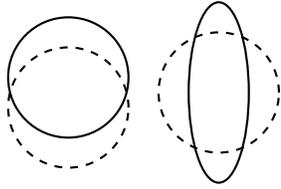
\begin{figure}
            \centering
            \begin{tikzpicture}[scale=1]
                \Large
                \node (O) at (0,0){};
                \draw[thick, dashed] ([yshift=-2mm]O) circle (8mm);
                \draw[thick] ([yshift=2mm]O) circle (8mm);
                \node (O2) at (2,0){};
                \draw[thick, dashed] (O2) circle (8mm);
                \draw[thick] (O2) ellipse (4mm and 12mm);
            \end{tikzpicture}
            \caption{Examples of dipolar and quadrupolar oscillations. Note that the dipolar case involves a shift in the center of mass. The quadrupolar case, in turn, encodes the deviations of the source from spherical symmetry.}
            \label{fig:vars}
        \end{figure}

        The quadrupole is the first moment containing information on the non-sphericity of the source. As such, it encodes the simplest possible deviations from a spherical distribution, such as the source's elongation or whether it possesses lobes. This opens up another possible way to understand why there is no gravitational dipole radiation. Let us look again at figure~\ref{fig:poles}, but focusing on the distribution of mass rather than charge. Clearly, varying the dipole means shifting the distribution of mass from one hemisphere to the other, which clearly shifts the position of the center-of-mass. From momentum conservation we know this never happens in the absence of external forces, so a gravitational dipole radiation does not exist. A quadrupole variation means, on the other hand, that part of the gravitational charge is moving from the equator to the poles, or vice-versa, in such a way that the center-of-mass remains fixed. These situations are illustrated in figure~\ref{fig:vars}.
                
        Interestingly, we can also see, from the representation of a pure quadrupole in figure~\ref{fig:poles}, that a rotation by $180^\circ$ brings the configuration back to its starting point. This means that a rotating quadrupole will emit radiation with twice the frequency of its rotation, as we saw in section~\ref{sec:binary}.
        
        \section{Why are  tensors ubiquitous in physics?}
        \label{sec:groups}
        
        The entire edifice of physics rests upon the fact that,   despite the many discrepancies in observations made by different individuals, there exist some relations among measured quantities that will be universally valid, and will be agreed upon by every rational entity in the Universe. For instance, even if an object may be perceived at rest for one observer, while for another it may be moving, both will agree on some laws dictating the dynamical causes and consequences of its motion. The main task of physics is to identify these relations and to write them in the form of universally valid statements. 
        
        A key step in this pursuit is to identify certain \emph{symmetries} of nature, i.e. sets of transformations that we can make on systems without affecting the validity of the physical laws.
        Once we write a physical law valid for one observer, it can be immediately translated to every other reference frame related to the first by such transformations. Some quantities may change in the process --- reflecting the differences seen by one observer as compared to another ---, but the relation among them as expressed by the physical law remains valid.
        
        This general requirement imposes stringent constraints on how the value of a physical quantity compares among different observers. Two observers may measure different values for a certain quantity, but these values must be related to one another by a specific rule determined by the underlying symmetry. And this in turn specifies the kind of mathematical  objects that must be used to describe physical quantities.
        
        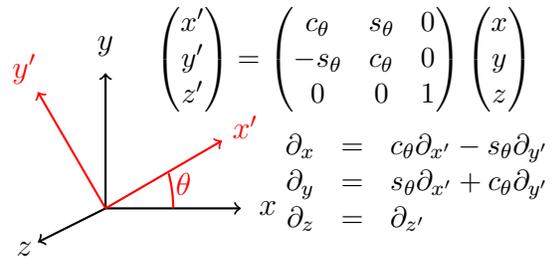
\begin{figure}
        \centering
        \begin{tikzpicture}[scale=1.8]
			\draw[->, thick] (0,0) --node[pos=1.2]{$x$} (1,0);
			\draw[->, thick] (0,0) --node[pos=1.2]{$y$} (0,1);
			\draw[->, thick] (0,0) --node[pos=1.2]{$z$} (-0.5, -0.25);
			\draw[->, thick, red] (0,0) --node[pos=1.2, red]{$x^\prime$} (0.86, 0.5);
			\draw[->, thick, red] (0,0) --node[pos=1.2, red]{$y^\prime$} (-0.5, 0.86);
			\draw[red, thick] (.5,0) arc (0:15:1) node[xshift=2mm, yshift=-4pt]{$\theta$};
			\node at (1.76,1.1) {
			    \small
				$\begin{pmatrix} x^\prime\\y^\prime\\ z^\prime \end{pmatrix} 
					= \begin{pmatrix} c_\theta& s_\theta & 0\\ -s_\theta & c_\theta & 0 \\ 0 & 0 & 1 \end{pmatrix} 
					\begin{pmatrix} x\\y\\z\end{pmatrix}$};
			\node at (2.3,0.25){
			\parbox{7cm}{\small\begin{eqnarray*}
				\partial_x &=& c_\theta \partial_{x^\prime} - s_\theta \partial_{y^\prime}\\[-1mm]
				\partial_y &=& s_\theta \partial_{x^\prime} + c_\theta\partial_{y^\prime}\\[-1mm]
				\partial_z &=& \partial_{z^\prime}
			\end{eqnarray*}
			}
		};
		\end{tikzpicture}
		\caption{Rotation of cartesian axes around the $z$-direction by an angle $\theta$, and the corresponding transformation rules for coordinates and derivatives.}
		\label{fig:frames}
		\end{figure}
		
		A typical example is the symmetry under rotations: because there are \emph{a priori} no preferred directions in space, any two frames rotated with respect to one another must observe the same physics. To be more concrete, consider a certain theory described by three fields $F_1,F_2,F_3$ which, in a certain reference frame $S$, obeys the equation
        \begin{equation}\begin{split}
            \partial_x F_1 + \partial_y F_2 + \partial_z F_3 &= 0.
            \label{eq:GaussS}
        \end{split}\end{equation}
        If we now look at the system from another frame $S^\prime$, related to $S$ via a rotation around the $z$-axis by an angle $\theta$, the description of spatial locations in $S^\prime$ will be different than in $S$, but both are related via the transformations shown in figure~\ref{fig:frames}.
        The derivatives will also be transformed accordingly, and equations~(\ref{eq:GaussS}) become
        \begin{equation}\begin{split}
           (c_\theta \partial_{x^\prime} - s_\theta \partial_{y^\prime}) {F_1} 
					+ (s_\theta \partial_{x^\prime} + c_\theta\partial_{y^\prime}){F_2} 
					+ \partial_{z^\prime}{F_3} &= 0.
				\label{eq:GaussSp}
        \end{split}\end{equation}
        
		It \emph{seems} that our physical laws have changed, that there is now an explicit dependence on $\theta$. But equations~(\ref{eq:GaussSp}) are still written in terms of the fields measured in $S$. We must now ask ourselves: if $S^\prime$ were to repeat the same procedures as $S$ to measure the fields, which values would it obtain? If the field values were the same in both frames --- i.e. if the fields are invariant under rotations --- then, based on this law, the observer in $S^\prime$ would be able to use these field measurements to empirically determine its rotation angle relative to $S$, contradicting the expectation that there is no preferred direction in space. Put another way, the directions of the axes in $S$ would be privileged because this law would assume a simplest form in this case. The situation can be remedied if (and only if) we demand the fields to also transform according to
		\begin{equation}
		    \begin{pmatrix} F_1^\prime\\F_2^\prime\\ F_3^\prime \end{pmatrix} 
					= \begin{pmatrix} c_\theta& s_\theta & 0\\ -s_\theta & c_\theta & 0 \\ 0 & 0 & 1 \end{pmatrix} 
					\begin{pmatrix} F_1\\F_2\\F_3\end{pmatrix}.
				\label{eq:rot_xy}
		\end{equation}
	 It is then easy to see that equation~(\ref{eq:GaussSp}) becomes
		\begin{equation}
		    \partial_{x^\prime}F_1^\prime +
		    \partial_{y^\prime}F_2^\prime
		    + \partial_{z^\prime}F_3^\prime=0,
		\end{equation}
		and the laws have the same form in both frames.
        
        \subsubsection*{Vectors} 
        
        Therefore the requirement that this law remain valid under an arbitrary rotation forces us to consider the three fields as three components of a larger object, which mix among themselves under such transformation. It is this larger object that is the truly invariant physical field. The transformation rule shown in equation~(\ref{eq:rot_xy}) can be more generally written as
        \begin{equation}
            F^\prime_i = \frac{\partial x^\prime_i}{\partial x_j} F_j,
            \label{eq:vec_trans}
        \end{equation}
        with Einstein's summation convention implicit. An object transforming according to this rule is called a \emph{vector}.
        
        Now suppose we add three more equations to the set of physical laws satisfied by these fields, describing how they are produced by three sources $J_1,J_2$ and $J_3$, namely
        \begin{equation}\begin{split}
            \partial_y F_3 - \partial_z F_2 &= J_1,\\
			\partial_z F_1 - \partial_x F_3 &= J_2,\\
            \partial_x F_2 - \partial_y F_1  &= J_3.
            \label{eq:Ampere_S}
        \end{split}\end{equation}
        If the source transforms as the vectors as well, these equations, as seen by $S^\prime$, would become
        \begin{widetext}\begin{equation}\begin{split}
			c_\theta (\partial_{y^\prime} {F_3^\prime}
			- \partial_{z^\prime} {F_2}^\prime)
			- s_\theta (\partial_{z^\prime} {F_1}^\prime
			-\partial_{x^\prime} {F_3}^\prime) &= c_\theta J_1^\prime - s_\theta J_2^\prime,\\
			s_\theta (\partial_{y^\prime} {F_3}^\prime - \partial_{z^\prime} {F_2}^\prime) 
			+ c_\theta (\partial_{z^\prime} {F_1}^\prime - \partial_{x^\prime} {F_3}^\prime)
			&= s_\theta J_1^\prime + c_\theta J_2^\prime
			,\\
			\partial_{x^\prime}  {F_2^\prime}
			- \partial_{y^\prime} {F_1^\prime}
					&= J_3^\prime.
        \end{split}\end{equation}\end{widetext}
        Here, something interesting happens: the equations are \emph{not} invariant under the transformation! However, in order for them to be simultaneously satisfied, a set of equations analogous to~(\ref{eq:Ampere_S}) would have to be obeyed in $S^\prime$ as well, i.e. replacing all unprimed variables by their corresponding primed versions. This then means that both $S$ and $S^\prime$ will again see the same physical laws. In this case the set of equations are said to be \emph{covariant} (rather than invariant). Covariance of the physical laws is really all we need to impose: it does not matter how one particular equation will transform from one frame to another, as long as the entire set is automatically satisfied in both frames.
        
        More generally, if we write a physical law as the vanishing of the components of a vector $\vec{V}$ in a frame $S$, as in $V_i=0$, then in another frame these equations become
        \begin{equation}
            \frac{\partial x_i}{\partial x^\prime_j} V^\prime_j =0,
        \end{equation}
        which implies (because the transformation matrix $\partial x_i/\partial x^\prime_j$ is invertible) that all components vanish in the frame $S^\prime$ as well. This is why mathematical objects like vectors, obeying a specific transformation rule as laid out in equation~(\ref{eq:vec_trans}), are the appropriate tools with which to formulate universal physical laws.
        
        \subsubsection*{Tensors}
        
        The argument presented at the end of the last subsection can be generalized in the following way. Let $T_{i_1\ldots i_n}$ be an object that transforms under some set of symmetry operation (e.g. rotations) as
        \begin{equation}
            T^{\,\prime}_{i_1i_2\ldots i_n} = \frac{\partial x^\prime_{i_1}}{\partial x_{j_1}}
            \frac{\partial x^\prime_{i_2}}{\partial x_{j_2}}\ldots \frac{\partial x^\prime_{i_n}}{\partial x_{j_n}} T_{j_1j_2\ldots j_n}.
            \label{eq:tensor}
        \end{equation}
        We call such an object a \emph{tensor of rank n}. Then, by the same arguments as above, any statement that these components vanish, $T_{i_1 i_2\ldots i_n}=0$, will be automatically satisfied in all frames related to each other by this particular symmetry operation (e.g. rotations). Put another way, any equation written in terms of objects which are tensors under a certain symmetry group will automatically be covariant under these transformations. This means that physical quantities will always be described by some kind of tensor under some set of symmetries. And we then know how each reference frame will compare their measurements to those made in another frame.
        
        Note that vectors are just rank 1 tensors. The quadrupole moment is an example of a rank 2 tensor: it has two axis that can be transformed independently, as discussed in appendix~\ref{sec:whytensor} above. Another particular type of quantities are the so-called \emph{scalars}, which are rank 0 tensors, i.e. objects which are invariant under said transformations.
        
        \subsubsection*{Four-vectors and covariant notation}
        
        Rotations are not the only kind of symmetry obeyed by physical laws. Since Galileo in the 16th century we know that all inertial frames experience the same physics, even those that are in relative uniform motion. Thus our physical objects must behave as tensors under transformations mapping differently moving inertial observers.
    
    But what is the form of these transformations? In order to understand this, note that the rotations discussed above are \emph{defined} as the transformations that keep spatial distances invariant, i.e. the quantity $\vec{dx}^2 = x^2+y^2+z^2$ is the same for every frame rotated with respect to each other. When we include relative movement among observers, the time coordinate obviously enters the picture as well, and we should then ask: which quantity related to time coordinates is invariant for every such observer? Galilean relativity attributes invariance to time intervals between events. But we now know this is not the case, since this would imply that different observers would measure different values for the speed of light, contradicting experimental results. Instead the truly invariant quantity is the speed of light $c$, so the invariant distance between two events is $ds^2 = c^2 dt^2 - \vec{dx}^2$. This means, in particular, that observers in relative motion will generally disagree on the length $\vec{dx}$ of an object or on time intervals $dt$, in precisely the way needed to ensure that both will measure $c=1$ for the speed of light. This is merely a statement that the propagation of light can act as a universal ruler for measuring distances between events, with which all inertial observers agree. The group of transformations that leave this quantity invariant is the so-called Lorentz group, and all physical quantities must be represented as tensors under Lorentz transformations, i.e. objects transforming according to the rule~(\ref{eq:tensor}). This guarantees that our physical laws will be valid for every inertial observer. These are called four-vectors and four-tensors.
    
    The situation can be put in geometrical terms as follows. The possibility of performing physical operations (namely rotations) that leave $\vec{dx}^2=x^2+y^2+z^2$ invariant leads us to understand these spatial coordinates as components of a larger three-dimensional object $(x,y,z)$ which mix among themselves when changing frames. Likewise, the invariance of $ds^2=c^2dt^2-\vec{dx}^2$ under all transformations among inertial observers leads us to the analogous understanding that time and space are also different dimensions of a larger four-dimensional entity which we call \emph{spacetime}. Note also that rotations are embedded in this larger class of transformations. Thus the usual vectors in three dimensions become just the spatial components of some larger four-dimensional objects. A point in spacetime, called an event, is then described by $x^\mu = (ct, \vec{x})$, with greek indices assuming values 0 (the ``time'' component) through 1, 2 and 3 (the ``spatial'' components of the four-vector).    Now the invariant measure of interval size can be written as $ds^2 = c^2dt^2 - |\vec{dx}|^2\equiv \eta_{\mu\nu} dx^\mu dx^\nu$ (recall that repeated indices are to be summed over), where $\eta_{\mu\nu} = \text{diag}(1,-1,-1,-1)$ is the so-called Minkowski spacetime metric. Different inertial frames are related to one another by some sort of ``rotations'' in spacetime that leave this metric invariant. In general, the inner product of two four-vectors can be defined as $v\cdot w \equiv \eta_{\mu\nu} v^\mu w^\nu$, and since this kind of combinations will often appear in calculations it is convenient to define a short-hand notation $v_\mu \equiv \eta_{\mu\nu} v^\nu$ for any four-vector $v$. Defining the inverse of the metric as $\eta^{\mu\nu}$, such that $\eta^{\mu\nu}\eta_{\nu \rho} = \delta^\mu_\rho$, it follows that $\eta^{\mu\nu} = \text{diag}(1,-1,-1,-1)$ and $v^\mu = \eta^{\mu\nu}v_\nu$. Thus the metric can be used to raise and lower indices, and in general raising and lowering indices with values $1,2$ and $3$ introduces a minus sign, while the $0$ component remains unchanged.
    
    Let $x^{\prime\,\mu}(x^\nu)$ represent a transformation between two intertial frames. Since it must preserve the spacetime interval we must have $ds^{\prime\,2}= \eta_{\mu\nu}dx^{\prime\,\mu}dx^{\prime\,\nu} 
     = 
     \eta_{\mu\nu} dx^{\mu} dx^{\nu}=ds^2$, so
     \begin{equation}
         \frac{\partial x^{\prime\,\mu}}{\partial x^\rho} \eta_{\mu\nu}
         \frac{\partial x^{\prime\,\nu}}{\partial x^\sigma} = \eta_{\rho\sigma}.
     \end{equation}
     This is the precise meaning of what we said above, that the Lorentz transformations leave the metric invariant. With this result, we find an easy way to construct scalars (i.e. invariant quantities with values agreed upon by all inertial frames) out of any tensor quantity: simply ``contract'' indices using the metric tensor until there are no free indices left. For example, out of the energy-momentum tensor $T^{\mu\nu}$ we can construct an invariant $T\equiv \eta_{\mu\nu}T^{\mu\nu}$ which is the trace of the tensor. This easiness to construct invariants out of tensors is a convenient consequence of writing our theory in terms of these objects.

    \subsubsection*{General relativity}
        
        The general theory of relativity is, as the name indicates, a generalization of this covariance principle to embrace all observers, even if non-inertial. In this case inertial frames are no longer special: every observer sees the same physics, and we must then write our objects as tensors under general coordinate transformations. This equivalence between inertial and non-inertial frames then means that, as we drop an object and observe it freely falling, we have no way of figuring out whether it fell because it was subjected to a gravitational field, or because of a fictitious force in an accelerating frame\footnote{This is analogous to the two inertial frames in relative motion who cannot agree on the electric and magnetic fields of a certain system.}. Both forces are in fact physically the same. Equivalently, this is to say that inertial mass is the same as gravitational mass. This is, in short, Einstein's equivalence principle.
        
        Since all frames are now equivalent, all physical laws should be expressed in terms of objects which behave as tensors (see eq.~(\ref{eq:tensor})) under \emph{general} coordinate transformations $x^\prime(x)$ --- including non-linear transformations leading to non-inertial frames. For instance, let us consider an infinitesimal coordinate transformation
        \begin{equation}
            x^{\prime\,\mu} = x^\mu + \xi^\mu,
        \end{equation}
        where the four-vector $\xi^\mu$ parametrizes the magnitude and direction of the transformation. The metric tensor transforms as
        \begin{equation}\begin{split}
            g^{\prime\,\mu\nu} &= \frac{\partial x^{\prime\,\mu}}{\partial x^\rho}\frac{\partial x^{\prime\,\nu}}{\partial x^\sigma}g^{\rho\sigma}\\
            &
            =(\delta^{\mu}_{~\rho} + \partial_\rho \xi^\mu)(\delta^{\nu}_{~\sigma}+\partial_\sigma \xi^\nu)g^{\rho\sigma}\\
            &=g^{\mu\nu} + \partial^\mu \xi^\nu + \partial^\nu \xi^\mu.
        \end{split}
        \label{eq:gaugegrav}
        \end{equation}
        With $g_{\mu\nu}=\eta_{\mu\nu} + h_{\mu\nu}$ and assuming $|h_{\mu\nu}|\ll 1$ (so we can keep only linear terms in the perturbation) the inverse metric is $g^{\mu\nu} = \eta^{\mu\nu}-h^{\mu\nu}$, and~(\ref{eq:gaugegrav}) then reduces to the gauge transformation stated in equation~(\ref{eq:gauge_h})\footnote{Note that, to linear order in $h_{\mu\nu}$, the raising and lowering of indices are performed by the Minkowski metric on both sides of the transformation equation.}.
        
        Finally, a side note: in quantum mechanics there also appears another type of object called a \emph{spinor}. The intuition behind this object is more elusive, but the core idea is similar to exposed above, namely that spinors are another type of objects that transform in a specific way under rotations or general Lorentz transformations. Just as tensors belong to a space generated out of ``product of vectors'', so does a vector also belong to a space generated by the product of two spinors. So spinors are called the \emph{fundamental} representation of the underlying symmetry group, since they generate every other transformation via tensor products.
        
        Everything we have discussed in this appendix refers to the branch of mathematics called \emph{group theory} and, more specifically, to \emph{representation theory} of groups. Groups are the appropriate abstract structures to describe transformations (as the symmetry operations we have been discussing), and representations are the objects (typically in a certain linear space) on which these transformations act in a specific manner, such as the tensors we have seen. Every physical object is described by a certain representation of the underlying symmetry groups of the theory, because physical objects must transform in a specific way in order to ensure the universality of the physical description among all observers. And it is important to understand how two objects in different representations combine to form a third one, e.g. how to combine two objects in order to form a scalar (i.e. an invariant) quantity, with whose value every observer agrees. This is why an understanding of representation theory is essential for the construction of physical theories out of symmetry arguments. Such a discussion is of course beyond the scope of this work, but the interested reader can consult references~\cite{Georgi, Gilmore} for a more in-depth discussion. An accessible reading on the subject of \emph{tensor calculus} for undergraduate students can be found in~\cite{Battaglia}. For a more geometrical approach, reference~\cite{Misner:1974qy} is recommended.

\end{document}